%% file: pension-paper-IME-format.tex
\documentclass{article}
\usepackage{authblk}
\usepackage{graphicx} % Required for inserting images
\usepackage{graphicx,epsfig}
\usepackage{amsfonts}
\usepackage{amsthm, amsmath}
\usepackage{amssymb}
\usepackage{subcaption}
\usepackage{xcolor}
\usepackage{textcomp}
\usepackage{listings}
\usepackage{multirow}
\usepackage{hyperref}
\usepackage{pdfpages}
\usepackage{enumitem}
\newtheorem{theorem}{Theorem}[section]
\newtheorem{proposition}[theorem]{Proposition}
\newtheorem{lemma}[theorem]{Lemma}

\newtheorem{definition}[theorem]{Definition}
\theoremstyle{definition}
\newtheorem*{notation}{Notation}
\usepackage{algorithm}
\usepackage{algpseudocode}
\usepackage{multirow,array}
\usepackage{booktabs}
\usepackage{appendix}
\usepackage{todonotes}
\usepackage[numbers]{natbib}

\newcommand{\RA}{\mathrm{RA}}
\newcommand{\BS}{\mathrm{BS}}
\newcommand{\ed}{\mathrm{d}}
\newcommand{\E}{\mathbb{E}}
\newcommand{\R}{\mathbb{R}}
\newcommand{\Q}{\mathbb{Q}}
\newcommand{\DT}{\ed {\cal T}(t)}

\newcommand{\id}{{\bf 1}}
\renewcommand{\P}{\mathbb{P}}

\DeclareMathOperator{\dom}{dom}

\DeclareMathOperator{\ri}{ri}

\DeclareMathOperator{\argmin}{argmin}

\newtheorem{remark}[theorem]{Remark}
\newtheorem{algoEnv}[theorem]{Algorithm}

\usepackage{bbm}

\title{Machine-learning a family of solutions to an optimal pension investment problem}
\author{
\small
John Armstrong\textsuperscript{a},
Cristin Buescu\textsuperscript{b},
James Dalby\textsuperscript{c},
Rohan Hobbs\textsuperscript{d,}\thanks{Corresponding Author. Strand, London, WC2R 2LS, UK. rohan.hobbs@kcl.ac.uk}\\
\small \textsuperscript{a,b,c,d}Department of Mathematics, King's College London, UK\\
\vspace{1em}
\small \textsuperscript{a}john.armstrong@kcl.ac.uk, 
\textsuperscript{b}cristin.buescu@kcl.ac.uk,
\textsuperscript{c}james.dalby@kcl.ac.uk,
\textsuperscript{d}rohan.hobbs@kcl.ac.uk
}

\date{}

\begin{document}

\maketitle
\begin{abstract}
We use a neural network to identify the optimal solutions
to a family of pension investment problems, where the parameters determining
an investor's risk and consumption preferences are given as inputs to the neural network in addition to economic variables. Training a single network across such a family fails without modification. Our main contribution is a scaling of the loss function that resolves this, together with a proof that the resulting algorithm converges. We use this to develop a practical tool for exploring how pension outcomes vary with preference parameters. We use a Black-Scholes economic model so that we may validate the accuracy of the network using a classical and provably convergent numerical method
developed using the duality approach.
\end{abstract}


\section*{Introduction}
The von Neumann--Morgenstern utility theorem implies that, under mild assumptions, an individual's preferences at a single time can be represented by an expected utility \cite{VNMTheorem}.
\input{pensions-introduction}

\input{pension-common-part1.tex}

\input{REVISIONS-pensions-common-part-2}

\section*{Acknowledgments}

JA and JD gratefully acknowledge funding from the Nuffield Foundation (grant FR-000024058). RH is funded by EPSRC DTP (grant ref: EPSRCDTP2401).

\bibliographystyle{unsrt}
\bibliography{upgrade}

\pagebreak

\appendix
\input{pension-appendix.tex}

\input{classical-method.tex}

\end{document}

%% file: pensions-introduction.tex
Nevertheless, even if one accepts that utility functions provide the best available tool to model preferences, it remains difficult to identify an
individual's preferences and this is a recurring criticism in the literature \cite{AfriatUtility, SamuelsonUtility, VarianUtility}.
When considering an investor's preferences over
time, the space of possible preferences is still larger. We
seek to provide a practical tool to assist
in identifying a reasonable approximation to an investor's preferences
for the purpose of pension investment. 

A standard practical approach taken when providing guidance on Defined Contribution (DC) investments is to 
give a
questionnaire to identify risk preferences \cite{Blake2020_PI2003}.
A pension professional designing products
can provide a menu of options to potential
customers and use tools such as a risk questionnaire to
advise them on the best selection from this menu. It
is in developing this menu of options that utility functions
can provide a useful framework. They allow preferences
to be operationalised mathematically and then used
to identify coherent investment strategies. Designs that are not based
on optimization may even prove to be stochastically dominated by other strategies
and this is clearly undesirable.

Our goal in this paper
is to build a tool a pension professional might use in order
to identify good candidate gain functions. It allows the pensions professional to
vary the parameters within a family of utility functions and quickly view the resulting outcomes. They can then
use their professional expertise to perform the 
subjective task of matching these outcomes to investors.
 Utility-function inference from behaviour has been studied mathematically \cite{cox2012utilitytheoryinferring, grzeskiewicz2025uncoveringutilityfunctionsobserved}, but we do not seek to do anything more sophisticated than
 inferring utility functions by selecting from a menu of options.

The principal difficulty in building such a tool is not the solution of any single optimal investment problem, but the simultaneous solution of a whole family of them, over parameter ranges wide enough to be useful in practice rather than over a small neighbourhood of a single calibration. The preference model we use is studied by machine-learning methods in \cite{armstrongDalbyHobbs1}, but there the parameters of the utility function are held fixed.

Learning a family of solutions with context is called amortized optimisation.
A naive approach is to generate random parameter values in addition to random economic scenarios and define the loss function as the expectation in the resulting product space. One can then minimize this loss function by stochastic gradient descent. This can work well when the objective function does not vary too greatly over the parameter space, for example in a hedging case when the certainty equivalent is used \cite{murray2022deephedging}, or in an optimal trading set up where the objective is linear quadratic \cite{Leal}.
However, the preference model we use contains an exponential term and as a result the gradients of the associated loss function differ by orders of magnitude across the parameter range. A small subset of parameter combinations then dominates the stochastic gradient descent, and the strategies learned for most parameter values are ineffective. As an alternative option, we also considered computing the loss function as a certainty equivalent cash value, but this is not analytically tractable for our gain function, and numerical root-finding brings its own difficulties when the scale of the problem is unknown.

Our main contribution is a technique that resolves this. We scale the loss function by a parameter-dependent factor, estimated online by auxiliary networks trained alongside the principal network, and chosen so that the variance of the stochastic gradients are equalised across parameter values. For each fixed set of parameters the scaling is an affine transformation of the objective, so the optimal strategies themselves are unchanged.

Similar normalisation techniques exist in machine learning, attaching a weight to each of a fixed list of objectives \cite{gradnorm} or tracking a single normalisation as it drifts over time \cite{popart}. Our problem is continuous in the parameters, so there is no finite list of objectives to weight. Nor is it enough to adapt over time, since the scale must vary across the parameters within a single run. Neither choice of scale is derived from a convergence guarantee, unlike our algorithm. There are also other gradient variance reduction techniques. Quasi-Monte Carlo, antithetic variates and importance sampling all reduce variance of an estimator for a fixed problem \cite{Glasserman}. Our
difficulty is not the precision of any individual gradient estimate. The scale of the loss varies by orders of magnitude across the preference family, so gradients computed at different draws of the preference parameters are incomparable however precisely each is estimated, and no amount of per-sample variance reduction equalises scales across the parameter space. These techniques are therefore complementary to our
scaling rather than alternatives to it, and may be applied alongside it.

We prove that the resulting scaled algorithm converges. Two results are required: the first is that the main iteration converges under any strictly positive, bounded, measurable scaling, and the second is that the scaling itself can be learned online.

To be precise, we provide such convergence results for an abstract stochastic gradient descent algorithm rather than for a full neural network algorithm. A convergence proof for a neural network algorithm can be broken down conceptually into the convergence of a stochastic gradient descent approximation and the approximation of an infinite-dimensional problem space by a network. The latter is usually the harder problem, but it is essentially orthogonal to the issues introduced by our scaling. Our results should therefore be read as an illustration of the modifications that any unscaled convergence proof would need in order to accommodate the technique. Convergence results for neural network approximations to stochastic optimal control problems do exist \cite{hure2021deep, han2016deep} and our arguments can be applied to them, but supplying the full details would obscure rather than highlight our contribution.

There is in any case a balance to be struck between provably convergent algorithms and effective implementation. Heuristic tools such as the Adam optimizer give good results even though a formal convergence proof may require a different rule for computing step sizes. Our technique can be applied whether or not the underlying algorithm is known to converge. Where it is known to converge, one expects the convergence proof can be modified as easily as we have modified that for stochastic gradient descent.

Our second contribution is to validate this numerical method for our preference model against a classical, provably convergent scheme obtained by a duality argument. This is why we work in a Black--Scholes market. It allows the accuracy of the network to be checked independently, and it keeps attention on the parameterisation of the utility function rather than on the economic model.

To produce an appropriate set of choices we first need a sufficiently rich family of gain functions to capture the key differences between different types of investor, while remaining easy to interpret. We require gain functions that capture preferences for consumption over time, while allowing individuals to distinguish between their risk-aversion and the diminishing marginal utility of consumption at any given point in time (which we call satiation). This distinction is necessary to resolve many asset pricing puzzles \cite{BansalYaron, Bansal, Benzoni}. Epstein--Zin preferences exhibit these features and offer analytically tractable solutions that can often be analysed with full rigour \cite{Herdegen1, Herdegen2, kraft2017optimal}. However, the positive homogeneity of such utility functions can produce some unrealistic solutions \cite{ArmstrongDalbyEpstein}. 
This motivates us to sacrifice analyticity and use a preference model given by Exponential Kihlstrom--Mirman preferences (also used in this context in \cite{armstrong2019collectivisedpensioninvestmentexponential}). Although general Kihlstrom--Mirman preferences
are not Markovian the exponential case without discounting that we use can be recast in recursive form \cite{EpsteinZin}. Bommier \cite{bommier2006} has argued from an axiomatic approach that this is a particularly appropriate model for problems featuring mortality risk. It also has the merit of making it easy to implement a forward looking machine-learning algorithm, while more general recursive preferences require backward algorithms which are considerably harder to implement and so are less accessible to practitioners.

The specific problem we solve is optimal investment with
idiosyncratic risk insured using a tontine structure. There is extensive literature on tontines; see \cite{Milevsky_2015} for a review of both the history of tontines and more recent literature. In\cite{DONNELLY201414} they describe a tontine mechanism to separate longevity risk from market risk that could be used in practice for this problem. We use an infinite fund approximation. Although tontines have been studied heavily, the literature on the optimal control approach to making best use of a tontine is surprisingly limited. The problem is solved for power utility in \cite{stamos2008optimal}, and
for Epstein--Zin utility with systematic longevity risk in \cite{armstrong_buescu_dalby}. 

The central
task in our paper, which is solving optimal control problems by machine learning, is well studied, see for example \cite{hure2021deep, han2016deep} and \cite{hu2023deep} for a recent survey of this fast-moving field.

Machine learning has also increasingly been applied to pensions, actuarial and insurance problems, particularly for pricing, risk assessment, and optimal control. Both \cite{GAN2013, HEJAZI2016169} consider variable annuity valuation, with the former using an ordinary Kriging method and the latter using neural networks for fast computation. Other studies have explored machine learning in insurance more broadly: in \cite{Jin2021} they propose a hybrid deep learning framework for optimal reinsurance, investment, and dividend strategies, while in \cite{Zhang2024} they reviewed Bayesian CART models for insurance pricing and risk modelling.

More directly relevant to our application, in \cite{LiForsyth2019}, they applied a data-driven neural network method to solve the multi-period optimal asset allocation problem during the accumulation phase of defined contribution pension plans. Reinforcement learning has also been applied to the optimization of pension portfolios \cite{irlam2020, Ozhamaratli2023}. Despite these works, there remains a gap in applying machine learning across the full pension lifecycle. We treat accumulation and decumulation together in a collective scheme and, unlike \cite{LiForsyth2019}, use an objective function that encompasses a richer family of preferences.

We have consciously chosen the most direct approach
of a forward method because we believe this will be the simplest for industry
to adopt without complex mathematics. Backward methods may be more efficient and scale better for higher-dimensional problems.

Section 1 describes the optimal pension investment problem at hand, and how we solve and validate this for fixed utility function parameter values. Section 2 provides detail on the scaling and our algorithm, including its convergence proofs. Section 3 discusses the results of our methods and Sections 4 \& 5 outline how our tool can be used for sensitivity analysis and the results of such work. We then conclude.

%% file: pension-common-part1.tex
\section{The Pension Problem and its Numerical Solution} \label{sec:the_problem}
\subsection{The Discrete-Time Investment Problem}
We consider an optimal investment and consumption problem set within a Black-Scholes framework. The dynamics of the risky asset price, denoted $S$, follow a geometric Brownian motion described by
\begin{equation*}
    \ed S_t = \mu S_t \ed t + \sigma S_t \, \ed W_t, \label{eq:BSM}
\end{equation*}
with $W$ a standard Brownian motion. Investment and consumption decisions are assumed to be made at discrete intervals, defined by the set ${\cal T}:=\{0,\delta t, 2 \delta t, \ldots, T \}$ for some time-step $\delta t$ and final time $T$. Between these times, investments are made following a fixed-weight strategy. Let $\pi_t$ be the proportion of wealth allocated to the risky asset at time $t \in {\cal T}$. This proportion is fixed throughout the interval $[t,t+\delta t)$ with the remainder growing at a constant rate $r$. The wealth evolves according to the stochastic differential equation (SDE)
\begin{equation}
    \ed w_s = w_s(\pi_t \mu + (1-\pi_t)r)\ed s + w_s \pi_t \sigma \, \ed W_s,\label{eq:returns}
\end{equation}
on the interval $[t, t+\delta t)$. We denote the limit from the left of wealth at the end of the period by $w_{(t+\delta t)-}$.
By It\^o's lemma, the log wealth process follows the linear process
\begin{equation*}
    \log(w_{(t+\delta t)-}) - \log(w_t) = \left(\pi_t \mu + (1-\pi_t)r-\frac{1}{2}(\pi_t\sigma)^2 \right) \delta t + \pi_t\sigma (W_{t+\delta t}-W_t).
\end{equation*}
In particular, the wealth is always positive.
Define
\begin{equation}
\epsilon_t := \frac{W_{t+\delta t}-W_t}{\sqrt{\delta t}}.
\label{eq:epsilon_as_normal}
\end{equation} 
We are able to simulate the log wealth process using the Gaussian increments $\epsilon_t$. 

To obtain $w_{t}$ from $w_{t-}$, we incorporate contributions, consumption and longevity payments via the equation
\begin{equation}
    w_{t} = \eta s_t \mathbbm{1}_{t<t_{\RA}} + (1- c_t
    \mathbbm{1}_{t\geq t_{\RA}})(1+ P_{\infty,t} \mathbbm{1}_{t>t_{\RA}}  )w_{t-},
\label{eq:wealthInfinite}    
\end{equation}
where the first term describes the fraction, $\eta$, of an individual's salary, $s_t$, that is contributed before retirement ($t_{\RA}$ is the time of retirement) and the second term removes the consumption, $c_tw_{t-}$ with $c_t \in (0, 1)$, and adds on any longevity payment, $P_{\infty, t}w_{t-}$, that one may receive in retirement. We choose to bound consumption at any time point so that $0 < \underbar{C} \leq c_t w_{t-} \leq \bar{C} < \infty$. The lower bound models a minimum subsistence level of consumption that any viable retirement strategy must provide, while the upper bound models regulatory drawdown limits on pension withdrawals. Since all of the terms in equation \eqref{eq:wealthInfinite} are positive, the wealth cannot become negative.

Our longevity credit, $P_{\infty, t}w_{t-}$, is derived from a tontine-style pooling mechanism (see \cite{armstrongDalbyHobbs1} for full implementation details). Each year, the funds of members who die are redistributed among survivors in proportion to their expected contribution to the pool. Crucially, redistribution is calculated after investment returns for the year are realised. This means members are not exposed to one another’s investment choices. In the general (finite) setting, the total funds released by those who die over the year are shared among surviving members in proportion to their odds of dying multiplied by their fund value:
\[
\frac{p^i}{1 - p^i} v^i,
\]
where $v^i$ is the realised end-of-period fund value for individual $i$ and $p^i$ is the corresponding probability of dying in the given year, conditional on survival to the previous year. In the special case of an infinite fund of identical individuals with no systematic longevity risk, the longevity payment simplifies to
\[
P_{\infty,t} w_{t-} = \left(\frac{p_t}{1 - p_t}\right) w_{t-},
\]
where $p_t$ is the probability of death during year $t$, conditional on survival to $t-1$, and $w_{t-}$ is the individual’s fund value immediately before the payment. The subscript $\infty$ references the fact that the fund is of infinite size, which we use for consistency of notation with our related work.

The optimization procedure is based on a stylized gain function which
we call Exponential Kihlstrom-Mirman utility \cite{armstrong2019collectivisedpensioninvestmentexponential}. These preferences are a variation on the concept of Kihlstrom–Mirman preferences \cite{KIHLSTROM1974361}, but additionally incorporate mortality. Specifically, the agents seek to maximise
\begin{equation}
    U(C):= \mathbb{E}\left( -\exp\left(-\alpha \sum_{j=t_{\text{RA}}}^{j<\tau} u(C_j) \delta t\right)\right),
    \label{eq:exponentialKMutility}
\end{equation}
where $\alpha > 0$ is a {\em risk aversion} parameter, $\tau$ is the individual's time of death and $\delta t$ defines the time step between consumption decisions. The \textit{felicity} $u(C_t)$ is given by
\begin{equation*}
    u(C_t):=\frac{C_t^\rho}{\rho}-\frac{a^\rho}{\rho},
\end{equation*}
where $\rho$ is a {\em satiation} parameter, $a$ is the {\em adequacy level} and $C_t$ is the individual's replacement ratio. The replacement ratio is defined to be the ratio of pension payments to index-linked final salary. Note that since overall consumption at any time point is bounded, the replacement ratio is too. Unlike Epstein--Zin preferences, this gain function involves no recursive computation of conditional expectations, so repeated evaluation is cheap.

Kihlstrom--Mirman preferences are not widely used in economic literature, but Bommier \cite{bommier2006} motivates them from an axiomatic approach, and argues that they are a strong candidate for problems with mortality. To obtain the general form of these preferences, one replaces the function $\exp$ in  \eqref{eq:exponentialKMutility} with an arbitrary concave function. As was observed by Epstein and Zin, in general, the resulting preferences are not Markovian \cite{EpsteinZin} and depend upon previous consumption. This need not in itself be a concern, but when there is discounting the dependence is implausible in that the impact of consumption history increases as the time since the consumption event increases. The one exception to both these issues is the case we consider of exponential preferences with no discounting. In this case, an induction argument shows that the preferences can be recast in the following recursive form where
they become manifestly Markovian:

\begin{equation}
V_s(C)=\begin{cases}
\alpha u(C_s)\,\delta t - \log \mathbb{E}\!\left[\exp(-V_{s+1}(C)) \,\middle|\, \mathcal{F}_{s}\right], & s<\tau,\\[2pt]
0, & s\ge\tau,
\end{cases}
\label{eq:markovianForm}
\end{equation}
so that
\begin{equation*}
U(C)=-\exp(-V_{t_\text{RA}}(C)).
\end{equation*}

Von Neumann-Morgenstern preferences are invariant under affine shifts and so the zero level of the felicity, $u$,  has no significance. This is not the case for these preferences. We have called the zero level of the felicity ``adequacy'' because living an extra year with an inadequate pension will reduce utility overall. 
We do not impose a hard constraint on adequacy, it is possible that an optimal pension strategy will risk some years of inadequate pension. In the von Neumann-Morgenstern case, the investor is entirely risk-neutral over adequacy, but as the risk aversion parameter $\alpha$ is increased, the risk-aversion increases. As we will see for very large $\alpha$ this will then determine the strategy. An alternative approach to modelling a notion of adequacy is to impose a hard constraint that consumption must never become inadequate. Our model instead provides a soft constraint for adequacy that is tied to risk aversion.

A popular alternative model for preferences, at least without mortality, is given by homogeneous Epstein--Zin preferences. These are positively homogeneous which leads to analytic tractability but at the cost of setting the adequacy level (as understood in terms of the felicity) to either $0$ or $\infty$. The moment one models adequacy as a positive real, this inevitably breaks positive homogeneity whether one uses a soft-constraint or a hard-constraint to model the concept.

Although we can Markovianise our preference model as in equation \eqref{eq:markovianForm}, we have not exploited this in the design of our neural network. This is intentional: a practitioner may wish to add some non Markovian features to the preference model such as further smoothing or loss aversion. We wish to demonstrate that the neural network approach we use does not depend upon Markovianity. 

In order to compute the gain function, we assume that consumption and individual longevity risk are independent. We also assume there is no systematic longevity risk and that the unconditional probability an individual dies in a given year $s$ is given by $\overline{p}_s$. So, we compute
\begin{equation*}
    U = -\mathbb{E}_{\mathrm{Invest}} \left[ \sum_{s=t_{\RA}}^T \overline{p}_s\, \text{exp}\left(-\alpha\sum_{t=t_{\RA}}^{s}  u(C_t) \delta t\right) \right].
\end{equation*}
In this formula, ${\mathbb E}_\mathrm{Invest}$ denotes the expectation across investment scenarios
and so excludes the mortality component of our probability model, which is accounted for by the term ${\overline{p}}_s$. $T$ is the maximum time of death, which is finite for the mortality model we are using. That is, all individuals are dead at time $T$.

In order to compute this loss function empirically, we generate $N$ investment scenarios and write $C^{(s)}$ for the replacement ratio in scenario $s$. Since neural networks minimise rather than maximise, we take as our objective the logarithm of the negative of the empirical gain function, giving the empirical loss
\begin{multline}\label{eq:loss}
    L = \log\left(\sum^N_{s=1}\exp\left(\log\left(\sum_{t=t_{\text{RA}}}^T \exp\left(\log(\overline{p}_t)-\alpha \sum^t_{j=t_{\text{RA}}}u(C_{j}^{(s)}) \delta t\right)\right)\right)\right)\\
    - \log(N),
\end{multline}
where the intermediate computations may be found in \cite{armstrongDalbyHobbs1}. We evaluate this expression using the logsumexp function to reduce excessive rounding errors and to ensure numerical stability.

\subsection{Training and Validating the Fixed-Parameter Network}
\label{sec:fixed_problem}
Throughout this subsection the gain function parameters $\alpha$, $\rho$ and $a$ are held fixed across all scenarios, so that what we solve is a single member of the family of problems rather than the family itself. We refer to the resulting network as the fixed-parameter network. Solving this optimization by classical techniques is computationally expensive, taking several minutes for the decumulation-only problem at a single fixed set of parameter values, and it becomes infeasible in richer economic models. It also requires considerable reprogramming whenever the economic model or the gain function is changed. We therefore solve the problem with a recurrent neural network. The inputs are the standard Gaussians $\epsilon_t$ of \eqref{eq:epsilon_as_normal} together with the time points, and the outputs at each time step are an investment proportion $\pi^\theta_t$ and a consumption proportion $c^\theta_t$, where the superscript $\theta$ records the dependence on the network parameters and so changes as the network is trained. Before retirement the consumption output is ignored. Taking these strategies over the whole time period $t \in [0,T]$ we evaluate the loss \eqref{eq:loss}. We use years as time points in our simulation, so $\delta t = 1$.

We use a gated recurrent unit rather than a feed-forward network. The problem is Markovian in wealth, but the network observes only the driving white noise and must reconstruct the wealth process from its past decisions, so the input-to-decision mapping is not Markovian. The architecture, given in Figure \ref{fig:RNN_architecture} of the appendix, is deliberately problem-agnostic rather than tuned to this particular problem and is chosen specifically so that it may be applicable to a wider range of problems that may not be Markovian.

To confirm that the trained neural network strategies are correct, we compare them against a provably convergent classical algorithm. In Appendix \ref{appendix:cvgtMethod} we use a duality approach to derive such an algorithm for the decumulation-only problem with discrete-time consumption and continuous-time investment, solving the problem by duality when consumption is restricted to one period and then recursing. In complete markets it is easier to prove rigorous results this way than for primal methods, which typically require restrictive growth conditions or more delicate arguments \cite{herdegen2021elementary}.

In order to test the accuracy of strategies for investment purposes, we compare certainty equivalent values. Across 100 parameter combinations sampled at random from the ranges of Table \ref{tab:param_ranges}, with one network trained per utility function parameter combination and no per-combination tuning, the mean error in the certainty equivalent was $0.70\%$ ($2.03\% \, s.d.$). The network is also far faster than the classical method, evaluating in seconds compared to the excess of five minutes the numerical scheme requires. For all evaluations in this paper we use 65536 scenarios and 96 annual time points, reduced to 56 time points in the decumulation-only cases.

% \begin{figure}[htbp]
%   \centering
%   \begin{subfigure}[b]{0.49\textwidth}
%     \centering
%     \includegraphics[width=\linewidth]{varying_parameter_paper/fixed_parameter_plot.png}
%     \caption{}
%     \label{fig:fixed_plot}
%   \end{subfigure}
%   \hfill
%   \begin{subfigure}[b]{0.49\textwidth}
%     \centering
%     \includegraphics[width=\linewidth]{varying_parameter_paper/neural_network_decumulation_vs_convergent.png}
%     \caption{}
%     \label{fig:nn_vs_cvgnt_method}
%   \end{subfigure}
%   \caption{In panel (\subref{fig:fixed_plot}) \textcolor{red}{we plot retirement outcomes (replacement ratio) deciles for the fixed-parameter trained neural network for the whole of life problem. In panel (\subref{fig:nn_vs_cvgnt_method}), we plot the decumulation-only accuracy of the fixed network to the provably convergent algorithm.}}
%   \label{fig:fixed_and_verification}
% \end{figure}

We were also able to validate our approach for the decumulation problem with power utility, discrete-time consumption and continuous-time investment using the analytic solution of \cite{armstrong_buescu_dalby}, Theorem 2.2. We have further tested that the method produces good strategies in richer economic models, but defer discussion of this to another paper.

None of this machinery extends to a family of gain functions. The naive way to let the risk aversion, satiation and adequacy parameters of \eqref{eq:loss} vary is to sample the three at random in each scenario, supply them to the RNN as additional inputs, and use them in the computation of the loss, so that the network trains on the whole family at once. Training this way performs so poorly that the expected utility from its strategies is practically incomparable to the optimal strategies given by the fixed parameter networks, for the reason set out in the introduction. Removing this obstruction is the subject of the rest of the paper.

%% file: REVISIONS-pensions-common-part-2.tex
\section{Scaling the Loss Function}
\subsection{Motivation for the Scaling}
\label{sec:motivation_scaling}
We remove the obstruction identified at the end of Section \ref{sec:the_problem} by scaling the loss function dependent on the gain function parameters. We begin by establishing that we are free to rescale.

Consider a filtered probability space $(\Omega^M, \mathcal{F}^M, (\mathcal{F}_t^M)_{t \in \mathcal{T}}, \mathbb{P}^M)$ describing the financial market randomness. Similarly, consider the filtered probability space $(\Omega^A, \mathcal{F}^A, (\mathcal{F}_t^A)_{t \in \mathcal{T}}, \mathbb{P}^A)$ for the parameters of the loss function. It contains the constant filtration $\mathcal{F}_t^A = \mathcal{F}^A$ for all $t$. This allows us to analyse the product space $\Omega = \Omega^M \times \Omega^A$, which we equip with the product measure $\mathbb{P} = \mathbb{P}^M \otimes \mathbb{P}^A$. Throughout, the parameter space is compact: $(\alpha, \rho, a)$ ranges over the closed bounded box of Table~\ref{tab:param_ranges}. By construction of the product measure, the market randomness is independent of the utility function parameters.
 
Our goal is essentially to minimise the negative of the expected utility. As such, in the fixed-parameter problem we are seeking to identify an element of
\begin{equation*}
    \argmin_{(c, \pi) \in \mathcal{A}} -\mathbb{E}\left( U(c) \right),
\end{equation*}
for the admissible control set $\mathcal{A}$. Our parametrised goal is to find an element of
\begin{equation}
    \argmin_{(c, \pi) \in \mathcal{A}^*} -\mathbb{E}\left(U(c) \mid \mathcal{F}^A \right),
    \label{eq:parametrised_goal}
\end{equation}
for the admissible control set $\mathcal{A}^*$ that depends on the parametrisation. By the linearity of the conditional expectation, we can make any $\mathcal{F}^A$-measurable positive affine transformation and equation \eqref{eq:parametrised_goal} will still have the same solutions. So to find this solution, we may consider
\begin{equation}
    \argmin_{(c, \pi) \in \mathcal{A}^*} - \mathbb{E}\left( \frac{U(c)}{\lambda} \right),
    \label{eq:scaled_expected_utility}
\end{equation}
where $\lambda$ is any strictly positive $\mathcal{F}^A$-measurable function. The details of the specific scale we choose and its importance are outlined in the following subsections, but the main point here is that we can scale the loss function, and we have considerable freedom to do so in our problem.
 
\subsection{Proving the Convergence of a Scaled Stochastic Gradient Descent Algorithm}
 
In order to prove that our scaled algorithm works, two things must be shown. The first, that the main iteration converges under any positive, bounded, measurable scale (Theorem~\ref{thm:SGD_with_scaling}), and second, that the scale itself can be learned online  (Theorem~\ref{thm:scaling_convergence}).
 
To motivate our proposed scaling, we first recall standard convergence results for SGD (see \cite{handbookconvergencetheorems}) in an unparametrised set up. We then prove that an analogous convergence result holds when the loss function is scaled. We first define the problem we are trying to minimise in terms of the empirical loss function, where the expectation in our utility function is taken over samples.
 
\begin{definition}[Minimisation Problem]
We seek to minimise a function $L : \Omega^M \times \mathbb{R}^d \to \mathbb{R}$ of the form
    \begin{equation}
        L(\omega^M, x) = \frac{1}{N} \sum_{i=1}^N L_i(\omega^M_i, x),
    \end{equation}
    with respect to $x$, where $L_i : \Omega^M \times \mathbb{R}^d \to \mathbb{R}$ for $\omega^M \in \Omega^M$. We assume $\argmin_x L \neq \emptyset$ and that each $L_i$ is bounded below. The minimiser of such a problem is denoted $x^*$.
\end{definition}
In our setting, $L$ corresponds to the loss in \eqref{eq:loss}, up to a logarithmic transformation introduced for numerical stability. This transformation does not affect the optimisation problem and we therefore state the results in full generality. The proof we follow in \cite{handbookconvergencetheorems} takes each $L_i$ to be the loss within a batch of a given size. In our case, without loss of generality, we assume that $L_i$ is the per-sample loss (mini-batches of size one) and therefore our overall loss $L$ is the average across samples. In our specific problem, one would take the per-sample loss as given by the negative exponential utility:
\begin{equation*}
    L_i := \sum_{t=t_{\text{RA}}}^T \exp\left(\log(\bar{p}_t) - \alpha \sum_{j=t_{\text{RA}}}^t u\left(C_j^{(i)}\right) \delta t \right).
\end{equation*}
The control $x$ represents the investment and consumption strategies that are used to compute the replacement ratio in the loss function and so in our set up the control depends on the scenarios themselves, i.e. $x:= x(\omega^M)$. These strategies $x$ are then used to compute the wealth process and consumption process and thus the replacement ratios in retirement. Since the replacement ratio is bounded, the admissible controls take values in a compact set.

For later use, we define the variance of a vector as
\begin{equation*}
    \mathbb{V}\left[X\right] := \mathbb{E}\left[|| X - \mathbb{E}\left[X\right] ||^2 \right].
\end{equation*}
We write $\mathrm{SD}\left[X \right] := \sqrt{\mathbb{V}\left[X \right]}$.
 
Throughout, we write $\mathcal{F}^k$ for the $\sigma$-algebra generated by the initialisation and by all samples of the gradient descent algorithm drawn up to and including iteration $k-1$. Conditional expectations $\mathbb{E}_k[\,\cdot\,] := \mathbb{E}[\,\cdot \mid \mathcal{F}^k]$ are taken with respect to this filtration.

Let us first define the classical SGD algorithm. Given an initialisation $x^0$, the classical SGD algorithm produces iterates
\begin{equation*}
    x^{k+1} = x^k - \gamma\, \nabla L_{i_k}\!\left(\omega^M_{i_k}, x^k\right),
\end{equation*}
where the gradient is taken with respect to $x$ and, at each iteration $k$, the index $i_k$ is drawn uniformly at random from $\{1, \ldots, N\}$, independently of $\mathcal{F}^k$.

From this point forward we require that the $L_i$ satisfy Assumptions 4.2 (convexity) and 4.3 (smoothness) in \cite{handbookconvergencetheorems}. In our case, since the exponential is convex and strictly increasing, its composition with a convex function is convex and similar standard results hold for the smoothness of such functions, particularly given that consumption is bounded. The smoothness assumption gives us the constant $\ell_{\text{max}}$, which is the largest smoothness constant (Lipschitz constant of the gradients) among the $L_i$, for $i=1, ..., N$, as defined in \cite{handbookconvergencetheorems}. 

Under these assumptions we have the following convergence result of the classical SGD algorithm.

\begin{theorem}[Convergence of SGD Algorithm]
    Consider a sequence $(x^k)_{k \in \mathbb{N}}$ generated by the SGD algorithm, with a constant step size (constant in $k$) satisfying $0 < \gamma \leq \frac{1}{4\ell_{\text{max}}}$. It follows that for every $K \geq 1$,
        \begin{equation}
            \mathbb{E}\left[L\left(\omega^M, \bar{x}^K \right) - \inf_x L(\omega^M, x) \right] \leq \frac{||x^0 - x^*||^2}{\gamma K} + 2 \gamma \sigma^*_L,
            \label{eq:original_sgd_bound}
        \end{equation}
        where $\bar{x}^K := \frac{1}{K}\sum_{k=0}^{K-1}x^k$ and $\sigma^*_L := \mathbb{V}\left[\nabla L_i(\omega^M_i, x^*) \right]$. 
    \label{thm:SGD}    
\end{theorem}
For a proof, see the proof of Theorem 5.3 from \cite{handbookconvergencetheorems}.
 
We have considered only constant step sizes because it is the simplest version and we wish to focus only on the implications of scaling. Non-constant step size SGD proofs can be addressed with the same arguments. 

Taking $\gamma = \frac{\gamma_0}{\sqrt{K}}$ forces the right hand side of the bound in \eqref{eq:original_sgd_bound} to be of order $\frac{1}{\sqrt{K}}$. The specific choice of $\gamma_0$ can guarantee a rate of convergence for the algorithm. In particular, by Corollary 5.6 of \cite{handbookconvergencetheorems}, we can guarantee that $\mathbb{E}[L(\bar{x}^K) - \inf_x L] \leq \epsilon$ provided we take 
\begin{align*}
    \gamma_0 = \min\left\{ \frac{1}{4\ell_\text{max}}, \frac{||x^0 - x^* ||}{\sqrt{2\sigma^*_L}} \right\}
\end{align*}
and
\begin{align}
    K \geq \left(||x^0 - x^*|| \sqrt{\sigma^*_L} + ||x^0 - x^*||^2 \ell_\text{max} \right)^2 \frac{16}{\epsilon^2}.
    \label{eq:unscaled_rate}
\end{align}

The speed of convergence is highly dependent on $\sigma^*_L$, the variance of the gradients, through the choice of step size. The optimal $\gamma_0$ shrinks as the variance grows. In our setting, the utility parameters are sampled as inputs to the neural network, and the gradient variance differs by orders of magnitude across parameter combinations. Each combination therefore has its own optimal step size, but the network is trained in a single run across all combinations at once, so only one step size can be chosen. Tuning it to the high-variance combinations slows convergence for all the others, while tuning it to the low-variance combinations destabilises training on the rest. Within our computational budget, neither compromise was acceptable in practice. What is needed is a step size that adapts to each parameter combination without abandoning the single training run. Scaling the sample-wise loss $L_i$ by a parameter-dependent factor achieves exactly this, since the scale multiplies the gradient and so acts as a parameter-dependent adjustment to the effective step size. The specific choice of scale is derived later in this section, where this argument is made precise, but first we show that applying any positive, bounded scaling in this way does not harm the convergence of the algorithm.
  
Theorem~\ref{thm:SGD_with_scaling} below shows that incorporating such scaling yields a convergence result analogous to Theorem~\ref{thm:SGD}, in which the moving scale enters the bound only through fixed constants. The scaling we want to use later in the paper requires online estimation, but error in this estimate does not stop the algorithm from converging. The scale only affects the speed of convergence.
 
Our problem is now to minimise the empirical loss
\begin{equation*}
    L(\omega^M, \omega^A, x) = \frac{1}{N} \sum_{i=1}^N L_i(\omega^M_i, \omega^A_i, x)
\end{equation*}
with $L_i : \Omega^M \times \Omega^A \times \mathbb{R}^d \to \mathbb{R}$ for $\omega^A \in \Omega^A$. Now, the control itself depends on the market scenario and the parameters, i.e. that $x := x(\omega^M, \omega^A)$. This allows us to define the scaled loss using a parameter-dependent scaling factor $\lambda(\omega^A)$:
\begin{align*}
    L_{\lambda}(\omega^M, \omega^A, x) :=
    \frac{1}{N}\sum_{i=1}^N L_{i, \lambda}(\omega^M_i, \omega^A_i, x),
\end{align*}
where
\begin{align*}
    L_{i, \lambda}\left(\omega^M_i, \omega^A_i, x \right) := \frac{L_i(\omega^M_i, \omega^A_i, x)}{\lambda(\omega^A_i)}.
\end{align*}
In our set-up, we also use SGD to find the scale, since it requires estimation online. As a consequence, we consider a sequence of scales in the following result. Throughout, the subscript $\lambda$ on the iterates $x^k_\lambda$ indicates generically that they are produced by the scaled algorithm; it does not refer to any single fixed scale. The scaled SGD update rule is therefore given by
\begin{equation*}
    x_\lambda^{k+1} = x_\lambda^k - \gamma \nabla L_{{i_k}, \lambda^k}(\omega^M_{i_k}, \omega^A_{i_k}, x_\lambda^k).
\end{equation*}
This leads us to the convergence result for the scaled SGD algorithm. 
\begin{theorem}[Convergence of Scaled SGD]
    Consider the sequence $(x^k_\lambda)_{k \in \mathbb{N}}$ generated by the SGD algorithm minimising the scaled loss, with a constant step size satisfying $0 < \gamma \leq \frac{\lambda_{\text{min}}}{4 \ell_{\text{max}}}$. Let $(\lambda^k)_{k \in \mathbb{N}}$ be a sequence of scale factors, each $\lambda^k = \lambda^k(\omega^A)$ measurable with respect to the information $\mathcal{F}^k$ available at iteration $k$, and satisfying
    \begin{equation*}
        0 < \lambda_{\text{min}} \leq \lambda^k(\omega^A) \leq \lambda_{\text{max}} < \infty, \quad \text{for all } k \geq 0 \text{ and all } \omega^A \in \Omega^A.
    \end{equation*}
    Then it follows that for all $K \geq 1$
    \begin{align}
        \mathbb{E}\left[L\left(\omega^M, \omega^A, \bar{x}^K_\lambda\right) - \inf_x L(\omega^M, \omega^A, x) \right] &\leq \lambda_{\text{max}}\left(\frac{||x^0_\lambda - x^*||^2}{\gamma K} + \frac{2 \gamma \sigma^*_L}{\lambda^2_{\text{min}}} \right)
        \label{eq:scaled_sgd_bound}
    \end{align}
    with $\bar{x}^K_\lambda := \frac{1}{K} \sum_{k=0}^{K-1}x^k_\lambda$ and $\sigma^*_L := \mathbb{V}\left[\nabla L_{i}\left(\omega^M_i, \omega^A_i, x^*\right)\right]$.
    \label{thm:SGD_with_scaling}
\end{theorem}
\begin{proof}
    For a proof, see Appendix \ref{appendix_scaled_sgd_proof}.
\end{proof}
The details of the proof essentially copy the proof of Theorem~\ref{thm:SGD}. The changes needed to accommodate the scale sequence $(\lambda^k)_{k\in \mathbb{N}}$
are minor, and ultimately this
enters the bound only through the fixed constants $\lambda_{\text{min}}$ and $\lambda_{\text{max}}$. The inequality at each step in the proof is bounded over the interval $[\lambda_{\text{min}}, \lambda_{\text{max}}]$, so it holds uniformly in $k$ irrespective of the trajectory taken by the scale. Moreover, because every scaled objective shares the same minimiser $x^*$, the iterates are compared throughout against one fixed expression $||x^k_\lambda - x^*||^2$. Consequently, the movement of $\lambda^k$ affects only the multiplicative constants, and hence the speed of convergence, never the fact of convergence, provided the scale remains within its fixed positive bounds. This is the precise sense in which finding $\lambda^*$ online cannot destabilise the main optimisation.
 
The same basic argument can be expected to hold whatever algorithm one uses. We have illustrated the idea for simple stochastic gradient descent, but the argument could easily be plugged into existing convergence proofs, for example for neural network algorithms.
 
\subsection{Identifying a Scaling}
For a fixed scale $\lambda^*$, we can rewrite Theorem \ref{thm:SGD_with_scaling} so that the bound in equation \eqref{eq:scaled_sgd_bound} is given by
\begin{equation}
    \mathbb{E}\left[L_{\lambda^*} (\omega^M, \omega^A, \bar{x}^K_{\lambda^*}) - \inf_x L_{\lambda^*}(\omega^M, \omega^A, x) \right] \leq \frac{\left|\left| x^0_{\lambda^*} - x^* \right|\right|^2}{K\gamma} + 2 \gamma \sigma^*_{L_{\lambda^*}} ,
    \label{eq:fixed_scale_bound}
\end{equation}
where $\sigma^*_{L_{\lambda^*}} := \mathbb{V}\left[ \nabla L_{i, \lambda^*}(\omega^M_i, \omega^A_i, x^*) \right]$.

An appropriate choice of $\gamma$ can once again guarantee a rate of convergence for the algorithm. 
The rate in \eqref{eq:unscaled_rate} reveals a structural problem in our setting when the problem is unscaled. Conditioning the bound of Theorem~\ref{thm:SGD} on the parameters and taking $\gamma = \gamma_0/\sqrt{K}$ gives
\begin{equation}
    \mathbb{E}\left[ L(\bar{x}^K) - \inf_x L \,\middle|\, \mathcal{F}^A \right] \leq \frac{1}{\sqrt{K}} \left[ \frac{||x^0-x^*||^2}{\gamma_0} + 2\,\sigma^*_L(\omega^A)\,\gamma_0 \right],
    \label{eq:unscaled_conditional_bound}
\end{equation}
where $\sigma^*_L(\omega^A) := \mathbb{V}[\nabla L_i(\omega^M_i, \omega^A_i, x^*) \mid \omega^A]$. If $\gamma_0 \leq \frac{1}{4\ell_\text{max}}$, which we will assume for simplicity from now on, the right-hand side is minimised by the step size
\begin{equation*}
    \gamma_0 = \frac{||x^0-x^*||}{\sqrt{2\,\sigma^*_L(\omega^A)}}.
\end{equation*}
This optimal step size depends on the parameter combination, but the step size must be fixed across all samples at each iteration for the entire training run. The cost of this restriction, and the gain from including a scaling, are made precise in the following result.

\begin{proposition}
\label{prop:cvgnce_rate}
Let $\sigma_{\max} := \sup_{\omega^A} \sigma^*_L(\omega^A)$, which we assume finite, and fix parameter dependent tolerances $\epsilon(\omega^A) > 0$, with $\epsilon_* := \inf_{\omega^A} \epsilon(\omega^A)$ assumed strictly positive. Then the following hold.
\begin{enumerate}[label=(\roman*)]
    \item For any single choice of $\gamma_0$, the guarantee $\mathbb{E}[L(\bar{x}^K) - \inf_x L \mid \mathcal{F}^A] \leq \epsilon(\omega^A)$ across all parameter combinations holds provided
    \begin{equation}
        K \geq \frac{1}{\epsilon_*^2} \left[ \frac{||x^0-x^*||^2}{\gamma_0} + 2\,\sigma_{\max}\,\gamma_0 \right]^2.
        \label{eq:unscaled_K}
    \end{equation}
    Minimising over $\gamma_0$, the best single step size, $\gamma_0 = ||x^0-x^*||/\sqrt{2\sigma_{\max}}$, yields the bound
    \begin{equation}
        K \geq \frac{8\,||x^0-x^*||^2\, \sigma_{\max}}{\epsilon_*^2}.
        \label{eq:unscaled_worst_case}
    \end{equation}
    \item For a scaled algorithm with scale $\lambda(\omega^A)$,
    \begin{equation}
        \mathbb{E}\left[ L(\bar{x}^K_\lambda) - \inf_x L \,\middle|\, \mathcal{F}^A \right] \leq \frac{1}{\sqrt{K}}\left[ \frac{\lambda(\omega^A)\,||x^0-x^*||^2}{\gamma_0} + \frac{2\,\sigma^*_L(\omega^A)\,\gamma_0}{\lambda(\omega^A)} \right],
        \label{eq:scaled_conditional_bound}
    \end{equation}
    and, up to a multiplicative constant, the right-hand side is minimised by
    \begin{equation}
        \lambda^*(\omega^A) = \sqrt{\sigma^*_L(\omega^A)} = \sqrt{\mathbb{V}\left[\nabla L_i(\omega^M_i, \omega^A_i, x^*) \mid \omega^A\right]},
        \label{eq:optimal_scale}
    \end{equation}
    the conditional standard deviation of the gradient at the optimum.
    \item Fixing the free constant in \eqref{eq:optimal_scale} so that $\sigma^*_{L_{\lambda^*}} = 1$, the choice $\gamma_0 = ||x^0-x^*||/\sqrt{2}$ (assuming that this is less than $\frac{\lambda_\text{min}}{4\ell_\text{max}}$) yields
    \begin{equation*}
        \mathbb{E}\left[L_{\lambda^*}(\bar{x}^K_{\lambda^*}) - \inf_x L_{\lambda^*} \,\middle|\, \mathcal{F}^A \right] \leq \epsilon(\omega^A) \quad \text{for all } \omega^A,
        \label{eq:scaled_uniform_K}
    \end{equation*}
    if
    \begin{equation*}
        K \geq \frac{8\,||x^0-x^*||^2}{\epsilon_*^2}.
    \end{equation*}
\end{enumerate}
    \label{prop:optimal_scale}
\end{proposition}
\begin{proof} For a proof, see Appendix \ref{appendix_cvgnce_rate}.
\end{proof}
The iteration count guaranteed by the theory for the unscaled algorithm is therefore dictated by the worst gradient variance across the parameter combinations. Moreover, choice of step size affects the entire family: with $\gamma_0$ tuned to $\sigma_{\max}$, the bound \eqref{eq:unscaled_conditional_bound} for every parameter combination is only $||x^0-x^*||\sqrt{2\sigma_{\max}}/\sqrt{K}$, so the bound is determined by the worst case rate of order $\sigma_\text{max}$, however small the variance for the parameter combination. In our setting, where the variances differ by orders of magnitude, this can render the algorithm ineffective for certain parameter combinations.

By including a scaling we adjust the step size for each parameter combination through a single schedule, the scale acting in \eqref{eq:scaled_conditional_bound} as a parameter-dependent effective step size $\gamma_0/\lambda(\omega^A)$. The iteration count is then uniform over the parameter family: in contrast with \eqref{eq:unscaled_worst_case}, no parameter combination pays for the variance of another. Since $\sigma^*_L(\omega^A)$ is not known in advance, $\lambda^*$ must be estimated online, and Theorem~\ref{thm:SGD_with_scaling} guarantees that error in this estimate affects only the constants in the bound and never the fact of convergence.

If choosing $\lambda^*$ came at no cost, then we would choose it to equal the standard deviation of the gradients at the optimum. For this choice to be admissible in Theorem~\ref{thm:SGD_with_scaling}, we assume that the conditional standard deviation is uniformly bounded above and bounded away from zero, so that $\lambda^* \in [\lambda_{\text{min}}, \lambda_{\text{max}}]$. We note that truncating the estimated scale onto this interval, as done in the implementation of our algorithm, guarantees this for the estimates $\lambda^k$. 
So in full, the overall loss under the optimal scaling is written
\begin{equation*}
    L^{\text{opt}}_{\lambda^*}(\omega^M, \omega^A, x) = \frac{1}{N} \sum_{i=1}^N \frac{L_i(\omega^M_i, \omega^A_i, x)}{\mathrm{SD}\left(\nabla L_i(x^*) | \,  \omega^A_i\right)}.
\end{equation*}
 
For our problem, using this method, it was possible within our computational budget to solve the optimal investment problem across a range of parameter values.
 
However, we are only seeking to tune the constant in any convergence result using our scaling and so it is important to consider the computational cost of computing the scaling itself. The obstacle is that computing this ideal scaling requires per-sample gradients, which loses the efficiency of the batched backward pass exploited by standard automatic differentiation. As a result, the cost (in time or memory) of the gradient computation grows with the batch size in addition to the dimension of the main RNN, and the algorithm does not scale efficiently as that dimension is increased. In practice, this overhead limits the effective training time and makes the method inefficient.
 
In the next section we will describe an alternative heuristically motivated method which scales linearly in the dimension of the main RNN. 
 
\subsection{An Alternative Heuristic Method}
 
Given the computational issues raised in the previous section, we propose a heuristic alternative algorithm for learning the optimal solution across different utility function parameters. It is computationally efficient, easy to implement, performs well in practice and the convergence remains guaranteed since the scale can again be bounded in a positive region. Importantly, this method does not require computing per-sample gradients.
 
Our heuristic approach is to scale the sample-wise loss by its own standard deviation, effectively normalising the variance of the loss function itself across parameter combinations. Although this approach does not directly minimise gradient variance, it serves as a computationally tractable proxy and performs well empirically.
 
The unparameterised version of our heuristic is commonplace in machine learning: it is standard practice to rescale data to have a mean of zero and standard deviation of one before fitting a neural network. Underlying this is the observation that for smooth problems, the natural scale of a function is similar to that of its derivatives. Informally, this last claim can be understood by noting that the smoothness of a function can be measured by the rate of decay of its Fourier coefficients.
 
In full, the heuristic method's loss function is given by
\begin{equation*}
    L^{\text{heur}}_{\lambda^*} (\omega^M, \omega^A, x) = \frac{1}{N} \sum_{i=1}^N \frac{L_i(\omega^M_i, \omega^A_i, x)}{\mathrm{SD}\left(L_i(x^*) | \, \omega^A_i \right)},
\end{equation*}
that is, the scaling is defined by
\begin{equation*}
    \lambda^*(\omega^A_i) = \mathrm{SD}\left[ L_i(\omega^M_i, \omega^A_i, x^*) \, | \, \omega^A_i \right].
\end{equation*}
\subsection{Online Conditional Moment Estimation}
Having established convergence under any admissible scale, we now describe how the
scale is estimated in practice.

Both scaling methods of the previous section depend on conditional moments evaluated
at the minimiser and at the sampled parameters $\omega^A$, the first of which is not
available during training and the second of which can only be known during the training procedure. We therefore estimate these moments online, at the current iterate
$x^k$, by a second SGD procedure running alongside the main optimisation. It has two
components: the first estimates a conditional mean, and the second uses that estimate
to form a conditional variance.

Write $Y_i$ for the quantity whose conditional variance is required, so that
\begin{equation*}
    Y_i = L_i(\omega_i^M,\omega_i^A,x^k) \in \mathbb{R}
    \qquad\text{or}\qquad
    Y_i = \nabla L_i(\omega_i^M,\omega_i^A,x^k) \in \mathbb{R}^d
\end{equation*}
in the heuristic and gradient-based methods respectively. The first component predicts
the conditional mean $\mu(\omega_i^A)$, valued in $\mathbb{R}^d$ in the gradient-based method, with one entry per component of the
gradient vector, and in $\mathbb{R}$ in the heuristic method. It minimises
\begin{equation}
    L_\mu(\mu, \omega^A)
    =
    \frac{1}{N}
    \sum_{i=1}^N
    \left\|
        Y_i - \mu(\omega_i^A)
    \right\|^2.
    \label{eq:L_mu}
\end{equation}
The second component predicts a scalar in both methods: by the definition we use, the variance of a random vector is the expected squared
Euclidean norm of its deviation from the mean, so the residual
$R_i := \| Y_i - \mu(\omega_i^A) \|^2$ is scalar in either case. It minimises
\begin{equation}
    L_v(v, \omega^A)
    =
    \frac{1}{N}
    \sum_{i=1}^N
    \left(
        R_i - v(\omega_i^A)
    \right)^2.
    \label{eq:L_v}
\end{equation}
The scale is then $\lambda(\omega^A) = \sqrt{v(\omega^A)}$. We truncate the scale and the mean at all iterations to a bounded set. This is required for the convergence of the algorithm. Since the parameter region is compact, we can find a large enough region so that the true mean and variance lie within the region.
 
We now prove convergence of the two-component SGD algorithm used to 
learn the scale. Rather than treating the two components separately, 
we prove the general result for a $J$-component cascade and recover 
our algorithm as the case $J = 2$.
 
The cascade is defined recursively. Setting $T_1^* := Y_i(x^*)$, each 
subsequent component regresses on the squared residual of the one 
before it,
\begin{equation*}
    T_j^* := \left|\left| T_{j-1}^* - m_{j-1}^* \right|\right|^2 \quad (j \geq 2), 
    \qquad 
    m_j^* := \mathbb{E}\left[T_j^* \mid \mathcal{F}^A\right] 
    \quad (j \geq 1),
\end{equation*}
so that $m_1^*$ is the conditional mean of $Y_i(x^*)$ and $m_2^*$ 
its conditional variance; higher components estimate the conditional 
variance of each preceding residual in turn. Our scaling uses only 
the conditional mean and conditional variance, but the following theorem covers the cascade 
at any depth. 
 
Each component minimises the least-squares objective 
$L_j(m) = \mathbb{E}[||T_j^* - m||^2]$, with 
$\nabla L_j(m) = 2(m - m_j^*)$, and is therefore $2$-strongly 
convex and $2$-smooth in the sense of 
\cite{handbookconvergencetheorems}. During training the true 
target is unavailable, so at iteration $k$ the implemented target 
$T_j^k$ replaces $Y_i(x^*)$ by the current main iterate $Y_i(x^k)$ 
and each $m_l^*$ by its current estimate $m_l^k$. Each component is updated using the implemented stochastic gradient $g_j^k := -2(T_j^k - m_j^k)$: the next iterate $m_j^{k+1}$ is the truncation of $m_j^k - \gamma g_j^k$ onto a closed bounded convex set $\mathcal{M}_j$. Such sets exist because each target $m_j^*$ depends continuously on the utility parameters, which range over a compact region, so the targets lie in a common bounded set and each $\mathcal{M}_j$ may be chosen to contain it.
Because the true target is unavailable, there is 
a bias on the gradient, and the argument below is an application of Theorem 5.8 of 
\cite{handbookconvergencetheorems} at the desired gradient of each component with an 
additional term accounting for that bias. 
\begin{notation}
For the theorem, we use the following notation:
\begin{align*}
e_j^k &:= \mathbb{E}\!\left[\lVert m_j^k-m_j^*\rVert^2\right],\\
E_x^k &:= \mathbb{E}\!\left[\lVert x^k-x^*\rVert^2\right],\\
\sigma_j^* &:= \mathbb{V}\!\left[-2(T_j^* - m_j^*)\right],\\
\Sigma^*_j &:= \max_{1\le l\le j}\sigma_l^*.
\end{align*}
 
Here, $e_j^k$ denotes the mean-square error of the $j$'th component estimate,
$E_x^k$ the mean-square error of the main iterate,
$\sigma_j^*$ the variance of the stochastic gradient of the $j$'th component at the optimum,
and $\Sigma^*_j$ the maximum gradient variance over all components up to component $j$.
\end{notation}
 
\begin{theorem}[Convergence of $J$-component SGD]
\label{thm:scaling_convergence}
Fix $J \geq 2$ and consider the $J$-component SGD cascade described above, in which 
the $j$-th component produces a sequence $(m_j^k)_{k \in \mathbb{N}}$ 
estimating $m_j^*$. Define the contraction rates $a_j := 2 - \frac{j-1}{J-1}$ and let
\begin{equation*}
    S_{x,(j)}^k := \sum_{s=0}^{k-1}(1 - a_j\gamma)^{k-1-s}E_x^s, 
    \quad 
    \Theta_j^0 := \sum_{l=1}^{j}\left|\left| m_l^0 - m_l^* 
    \right|\right|^2, 
    \quad 
    \Pi := 1 + J(J-1)C_B.
\end{equation*}
Suppose that:
\begin{enumerate}
    \item $Y_i(x)$ is Lipschitz in $x$ with constant $C_Y > 0$;
    \item $Y_i(x^*)$ has finite moments up to order $2^J$, that is 
    $\mathbb{E}\left[\left|\left|Y_i(x^*)\right|\right|^{2^J}\right] 
    < \infty$; in particular $\sigma_j^* < \infty$ for every $j$;
    \item all components share a constant step size $\gamma$ with 
    $0 < \gamma \leq \frac{1}{16}$.
\end{enumerate}
Then for every $j = 1, \ldots, J$ and all $k \geq 1$,
\begin{equation}
    e_j^k \leq \Pi^{j-1}\left[(1 - a_j\gamma)^k\,\Theta_j^0 
    + 4\gamma\,\Sigma_j^* + JC_B\,\gamma\,S_{x,(j)}^k\right].
    \label{eq:bound_on_scale}
\end{equation}
\end{theorem}
\begin{proof}
    See Appendix~\ref{appendix_scaling_convergence}.
\end{proof}
 
\begin{remark}[The bound vanishes to a controllable floor]
\label{rem:scaling_bound_to_zero}
If the main iterate converges in mean square, that is $E_x^k \to 0$ as $k \to \infty$, then $S_{x,(j)}^k \to 0$ and the 
right-hand side of \eqref{eq:bound_on_scale} converges to a noise 
floor of order $\gamma$. This floor is not a fundamental 
limitation: by Corollary 5.9 of \cite{handbookconvergencetheorems}, 
for any precision $\varepsilon > 0$ one can choose $\gamma$ and a 
finite number of iterations $K$ with $e_j^K \leq \varepsilon$.

The assumption that $x^k \to x^*$ in mean square is not a demanding constraint. Any strongly convex problem has this property (see the convergence results for strongly convex objectives in \cite{handbookconvergencetheorems}), provided the step sizes are chosen appropriately. We can guarantee this in the case of our main optimisation problem by the bounds on consumption. Strong convexity can be enforced more generally if it does not already hold, for example by 
Tikhonov (Ridge) regularisation.
\end{remark}

\begin{remark}[Lipschitz Assumption]
Assumption 1 in Theorem \ref{thm:scaling_convergence} is immediate for the gradient method from the smoothness assumption. In the heuristic method $Y_i = L_i$ and it is 
a mild additional requirement, holding for instance when the 
gradients are uniformly bounded.
\end{remark}

Our specific algorithm is the case $J = 2$, estimating the conditional mean 
$\mu^* := m_1^*$ and conditional variance $v^* := m_2^*$ of 
$Y_i(x^*)$. Here the cascade requires only 
$\mathbb{E}[\|Y_i(x^*)\|^4] < \infty$, and the theorem gives
\begin{equation*}
    \mathbb{E}\left[\left|\left| v^k - v^* \right|\right|^2\right] 
    \leq (1 + 2C_B)\left[(1-\gamma)^k\,\Theta^0 
    + 4\gamma\,\Sigma^* + 2C_B\,\gamma\,S_{x,(2)}^k\right],
\end{equation*}
where $\Theta^0 := ||\mu^0 - \mu^* ||^2 + ||v^0 - v^*||^2$ and $\Sigma^* = \max(\sigma^*_{L_\mu}, \sigma^*_{L_v})$, in which $\sigma^*_{L_\mu}$ and $\sigma^*_{L_v}$ denote the stochastic gradient variances at the optimum of the mean-component loss \eqref{eq:L_mu} and the variance-component loss \eqref{eq:L_v} respectively.
The scale is recovered as $\lambda^k = \sqrt{v^k}$ and since $\lambda$ is bounded above and below, the variance bound 
transfers directly to the scale:
\begin{equation*}
    \mathbb{E}\left[\left|\left| \lambda^k - \lambda^* 
    \right|\right|^2\right] \leq \frac{1}{4\lambda_{\min}^2}\,
    \mathbb{E}\left[\left|\left| v^k - v^* \right|\right|^2\right],
\end{equation*}
which can be made arbitrarily small by Remark~\ref{rem:scaling_bound_to_zero}. 
 
\subsection{Implementation Using Neural Networks}
In order to implement the joint optimisation of the main network and the two-component scaling procedure described in the previous subsection, we use three separate neural networks, one for each component of the optimisation. We have the standard main RNN which is the same as in the fixed-parameter set up, except that now it optimises the scaled loss function and takes the utility function parameters as inputs. Its convergence comes from Theorem \ref{thm:SGD_with_scaling}. The scaling is produced by two extra networks: we call the neural network optimising the first component of the scaling optimisation the `mean network', and we label the neural network optimising the second component the `scaling network', since this is what produces the scale that is used in the loss function. Their convergence comes from Theorem \ref{thm:scaling_convergence}.

The joint convergence comes from observing that there is no interdependence in the coupled procedure. Theorem~\ref{thm:SGD_with_scaling} requires only that the scale remain within fixed positive bounds, not that it be accurate, while Theorem~\ref{thm:scaling_convergence} requires only mean-square convergence of the main iterate. We have shown that 
the main control converges under any positive, bounded, 
$\mathcal{F}^A$-measurable scale, and that the scale our algorithm 
requires can itself be learned online by the two-component mechanism 
on a moving target. Together these establish convergence of the 
full procedure, with the scaling serving to reduce the gradient 
variance and thereby accelerate convergence to the solution.

Following our algorithm, the training procedure jointly optimizes all three networks using the losses defined in \eqref{eq:L_mu} and \eqref{eq:L_v}, as well as the main RNN's scaled loss. At each step of training $k$ (each gradient update), all three networks make a prediction given the dataset and parameters. The main RNN's strategy $(x^k_t)_{t=1}^T$, along with the scale from the scaling network, $\lambda^k(\omega^A)$, are used as input into the main RNN's (scaled) loss function $L_{\lambda^k}$. The two auxiliary networks' losses are computed using the (unscaled) sample-wise loss $L_i(\omega^M, \omega^A, x^k)$ from the main RNN, as per the loss functions $L_\mu(\omega^A)$ and $L_v(\omega^A)$ respectively. Of course the scaling network requires the prediction from the mean network to compute its loss function. Finally, all three networks' parameters are updated using the corresponding gradients from their loss functions. The full training procedure is given in Algorithm \ref{alg:joint_training}:

\begin{algorithm}[htbp]
\caption{Joint Optimization Training Procedure}
\label{alg:joint_training}
\begin{algorithmic}[0]
\State \textbf{Initialize:} market dataset $\Omega^M$, utility function parameters $\Omega^A$, network parameters $\theta_{\text{RNN}}, \theta_\mu,  \theta_v$, initialize iteration counter $k=0$, number of epochs $E$ and number of batches $B$
\For{epoch in range $E$}
    \For{batch in range $B$}
        \begin{enumerate}[label=\arabic*., leftmargin=4em, nosep]
            \item Sample scenarios $\omega^M \in \Omega^M$ and parameters $\omega^A \in \Omega^A$
            \item $(x_t^k)_{t=1}^T \gets \text{RNN}(\omega^M, \omega^A)$
            \item $\mu^k_\theta(\omega^A) \gets \text{MeanNN}(\omega^A)$
            \item $v^k_\theta(\omega^A) \gets \text{ScalingNN}(\omega^A)$
            \item Truncate $\mu^k_\theta(\omega^A)$ and $v^k_\theta(\omega^A)$ onto a bounded set containing $\mu^*$ and $v^*$.
            \item Compute state process from $(x_t^k)_{t=1}^T$ to get the outcome variable, $(C_t^k)_{t=1}^T$, that is used to compute loss function.
            \item Input $(C_t^k)_{t=1}^T$ and $\lambda^k(\omega^A):=\sqrt{v^k(\omega^A)}$ into the scaled RNN loss $L_{\lambda^k}$, and store appropriate variable $Y_i$ associated to each $L_i$ for all $i$ in the batch (either per-sample gradient or per-sample loss). 
            \item Compute mean-estimating network loss $L_\mu(\omega^A)$ using the stored $(Y_i)_{i=1}^N$, and store the residuals $(R_i)_{i=1}^N$ from this computation.
            \item Compute scaling network loss $L_v(\omega^A)$ using $(R_i)_{i=1}^N$
            \item Update $\theta_{\text{RNN}}, \theta_\mu, \theta_v$
            \item Increase iteration counter $k$ by one
        \end{enumerate}
    \EndFor
\EndFor
\end{algorithmic}
\end{algorithm}

In order to maintain numerical stability, we transform all of our loss functions into log space, and use numerically stable functions like \emph{logsumexp}. For further details, see Appendix \ref{appendix_log_transformations}.

\section{Accuracy of Our Methods}
\label{sec:scaling_results}
We tested the accuracy of our methods against fixed parameter neural networks for a range of 100 randomly sampled parameter combinations. The mean certainty equivalent percentage error was $4.28\%$ ($5.31\% \, s.d.$) for the gradient-based method and $0.284\%$ ($0.48\% \, s.d.$) for the heuristic method, across the 100 samples. This level of accuracy provides us with confidence that both of our algorithms have worked successfully, since we are comparing the results with fixed-parameter networks which we know to be highly accurate to a provably convergent solution in the decumulation only case. We have no reason to believe the accuracy deteriorates in the whole-of-life problem.

Whilst both methods are a marked improvement on the naive method of training, the heuristic method consistently outperforms the gradient-based method across different parameter combinations. The gradient-based method's accuracy is not as high as the heuristic method, and it does a worse job at generalising across parameters. This is because it cannot be trained as well as the heuristic method under the same computational capacity. Both methods were trained for 24 hours, after which the best-performing model from each run was selected for evaluation. However, the gradient-based method completed only approximately half as many training iterations and was limited to a batch size 32 times smaller. Hence it received two orders of magnitude less training.

Although the gradient-based approach may possess stronger theoretical foundations, its computational demands make it considerably less practical. In contrast, while the heuristic method provides no formal optimality guarantee for the rate of convergence, it achieves superior empirical performance.

% \begin{figure}[htbp]
%   \centering
%   \begin{subfigure}[b]{0.49\textwidth}
%     \includegraphics[width=\linewidth]{varying_parameter_paper/gradient_based_vs_fixed_rr.png}
%     \caption{}
%     \label{fig:gradient_based}
%   \end{subfigure}
%   \hfill
%   \begin{subfigure}{0.49\textwidth}
%     \centering
%     \includegraphics[width=\linewidth]{varying_parameter_paper/heuristic_vs_fixed_rr.png}
%     \caption{}
%     \label{fig:heuristic}
%   \end{subfigure}
%   \caption{Panel (\subref{fig:gradient_based}) shows retirement outcomes for the RNN produced by the gradient-based method as compared to a fixed-parameter network. Panel (\subref{fig:heuristic}) shows the same comparison but between the RNN from the heuristic algorithm and the same fixed parameter network. Each line on the graph is a decile of the replacement ratio from all scenarios.}
%   \label{fig:varying_networks_vs_fixed}
% \end{figure}

For a visual comparison of the accuracy of the methods for one parameter combination evaluation, see the plots in Figure \ref{fig:comparisons_vs_fixed_rrs} in Appendix \ref{AppendixPlots}.

\section{Real-Time Comparisons}
The primary objective of training this varying-parameter network was to enable real-time comparisons of retirement outcomes, allowing users to explore and select their preferred investment and consumption strategies. From here onwards, we will take the varying-parameter network to be the network produced by the heuristic method of training, since it had the best results.

Although the varying-parameter networks generate predictions over the full parameter range without requiring further training,  it remains time-consuming to simulate sufficient strategy outcomes to produce a fan diagram of outcomes. To address this issue, we train an additional feed-forward neural network, referred to as the `replacement-ratio percentile' network. This network approximates the mapping from percentile, input parameters, and time point to the corresponding replacement ratio, as predicted by the principal RNN. Specifically, let $\mathcal{I} := \{1, 2, \ldots, 9\}$ denote the set of deciles and let $\mathcal{T}^\text{ret}$ be the time points restricted to retirement times only. Then the replacement-ratio percentile network learns the mapping 
\begin{equation*}
    g : \Omega^A \times \mathcal{I} \times \mathcal{T}^\text{ret} \to \mathbb{R},
\end{equation*}
which returns the value of the $i$-th percentile at time $t$ in retirement, corresponding to the replacement ratio produced by the optimal strategy obtained from the principal network. The training was performed in a supervised environment and the dataset was generated by randomly sampling $50,000$ combinations of parameters and passing them through the principal RNN to obtain the corresponding nine percentiles for each input. Each percentile contains $56$ time points to account for each year from retirement until the last time point an individual may still be alive.

% We can see in Figure \ref{fig:graph_plot} the accuracy of this network in mimicking the principal network. 
% \begin{figure}[htp!]
%   \centering
%   \begin{minipage}[b]{0.80\textwidth}
%     \centering
%     \includegraphics[width=1.0\linewidth]{varying_parameter_paper/graph_network_plot.png}
%   \end{minipage}
%     \caption{Retirement outcomes for the replacement-ratio percentile neural network compared to the main RNN.}
%     \label{fig:graph_plot}    
% \end{figure}

As one would expect, the replacement-ratio percentile network produces an almost identical output to the principal RNN across the whole range of parameters. The replacement-ratio percentile network is also able to produce the plot in approximately 1/3 of a second, more than 10 times faster than recalculating by Monte Carlo. This allowed us to create
a far more interactive and responsive user interface.

\section{Sensitivity Analysis of Optimal Strategies} \label{sec:sensitivity_analysis}
The ability to learn the optimal strategy across a range of parameter values allows us to perform sensitivity analysis with respect to the parameters. Unless otherwise stated, all parameters are kept constant as specified in Table \ref{tab:parameters}. We will analyse the outcomes in retirement and the strategies themselves. The plots for outcomes will be included in this section, but the full plots of the strategies will be included in Appendix \ref{AppendixPlots}.
\medskip

We begin by examining how retirement outcomes vary with the risk-aversion parameter, $\alpha$. Figures \ref{fig:high_alpha_plot} \& \ref{fig:low_alpha_plot} below illustrate how the outcomes differ relative to this parameter.

\begin{figure}[htbp]
  \centering
  \begin{subfigure}[b]{0.49\textwidth}
    \includegraphics[width=\linewidth]{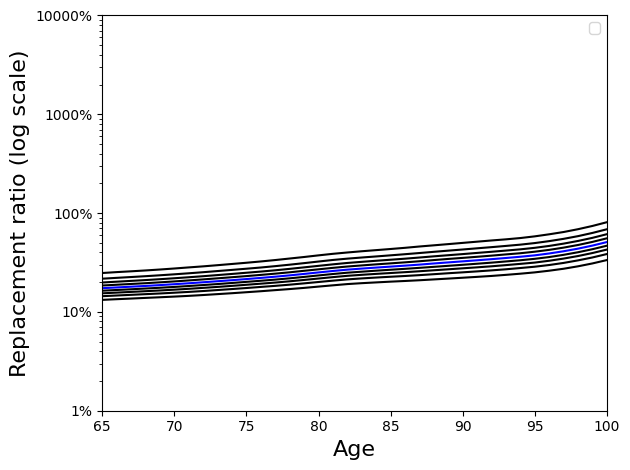}
    \caption{}
    \label{fig:high_alpha_plot}
  \end{subfigure}
  \hfill
  \begin{subfigure}{0.49\textwidth}
    \centering
    \includegraphics[width=\linewidth]{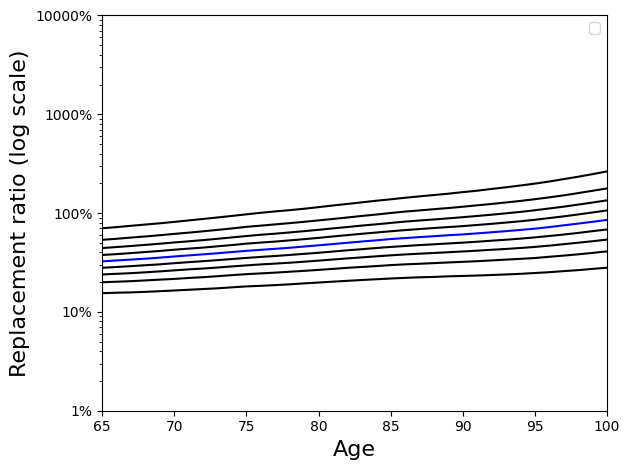}
    \caption{}
    \label{fig:low_alpha_plot}
  \end{subfigure}
  \caption{Comparison of retirement outcomes for different values of $\alpha$. We have subfigure (\subref{fig:high_alpha_plot}) with $\alpha = 0.01$ (high risk aversion) and subfigure (\subref{fig:low_alpha_plot}) with $\alpha = 10^{-7}$ (low risk aversion). Each line represents a decile from the simulation.}
  \label{fig:alpha_comparison}
\end{figure}
The impact of the risk aversion parameter is as expected. More risk averse individuals make less risky investments and consume less at the start of retirement in order to ensure they do not deplete their funds before dying. The opposite is true when the individual is less risk averse, and is happier to take on more investment risk and chance that their funds may deplete before death.
\medskip

\begin{figure}[htbp]
  \centering
  \begin{subfigure}[b]{0.49\textwidth}
    \includegraphics[width=\linewidth]{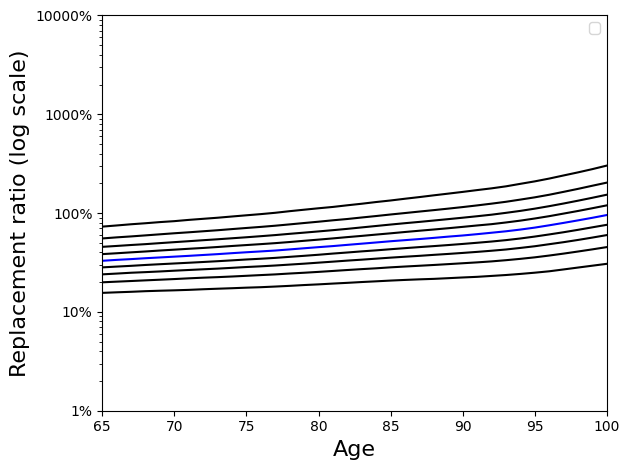}
    \caption{}
    \label{fig:high_rho_plot}
  \end{subfigure}
  \hfill
  \begin{subfigure}{0.49\textwidth}
    \centering
    \includegraphics[width=\linewidth]{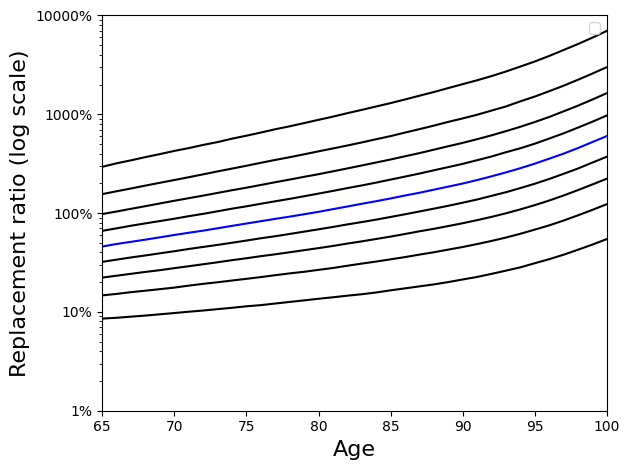}
    \caption{}
    \label{fig:low_rho_plot}
  \end{subfigure}
  \caption{Comparison of retirement outcomes for different values of $\rho$. We have Subfigure (\subref{fig:high_rho_plot}) with $\rho = -2.0$ (more easily satiated) and Subfigure (\subref{fig:low_rho_plot}) with $\rho = -0.1$ (less easily satiated). Each line represents a decile from the simulation.}
  \label{fig:rho_comparison}
\end{figure}

We can also consider the effect the satiation parameter, $\rho$, has on retirement outcomes. Figures \ref{fig:high_rho_plot} \& \ref{fig:low_rho_plot} provide an illustration. 

Again, the figures show that the outcomes follow an intuitive pattern. When an individual becomes satiated more quickly, they reduce investments in the risky asset, and consume less in retirement because they are satisfied with a lower consumption. The opposite is true when the individual becomes satiated less quickly and seeks to consume more in retirement.
\medskip 

A key point from the sensitivity analysis of these two parameters is that it is hard to distinguish between them in this economic model in the whole of life set up. This may be different in a model of stochastic interest rates, as one would expect to see the satiation parameter (the elasticity of inter-temporal substitution) tied closely to the state of the interest rate.

\medskip

The impact of the adequacy parameter itself is much harder to see than the other two parameters. To understand its effect, we examine the behaviour with extremely high risk aversion $(\alpha=0.2)$. Let $V^{\text{adequate}}$ be the level of funds at retirement that allows an individual to consume at the adequacy level for the whole of their retirement if they invest in risk-free bonds. If a highly risk-averse individual has more than $V^{\text{adequate}}$ at retirement, their strategy will approximate consuming all funds above $V^{\text{adequate}}$ in the first year and consume at the adequacy level thereafter. This is because this is the only risk-free strategy available. If they have less than $V^{\text{adequate}}$ funds, the situation is reversed. An individual who is risk averse will choose to reduce consumption initially below the level they could sustain for the whole of retirement in order to reduce their risk levels later in retirement.
This is illustrated in 
 Figure \ref{fig:cpp_risk_adequacy_plots}, where we deploy the provably convergent numerical scheme outlined in Appendix \ref{appendix:cvgtMethod}. 

\begin{figure}[htbp]
  \centering
  \begin{subfigure}[b]{0.49\textwidth}
    \includegraphics[width=\linewidth]{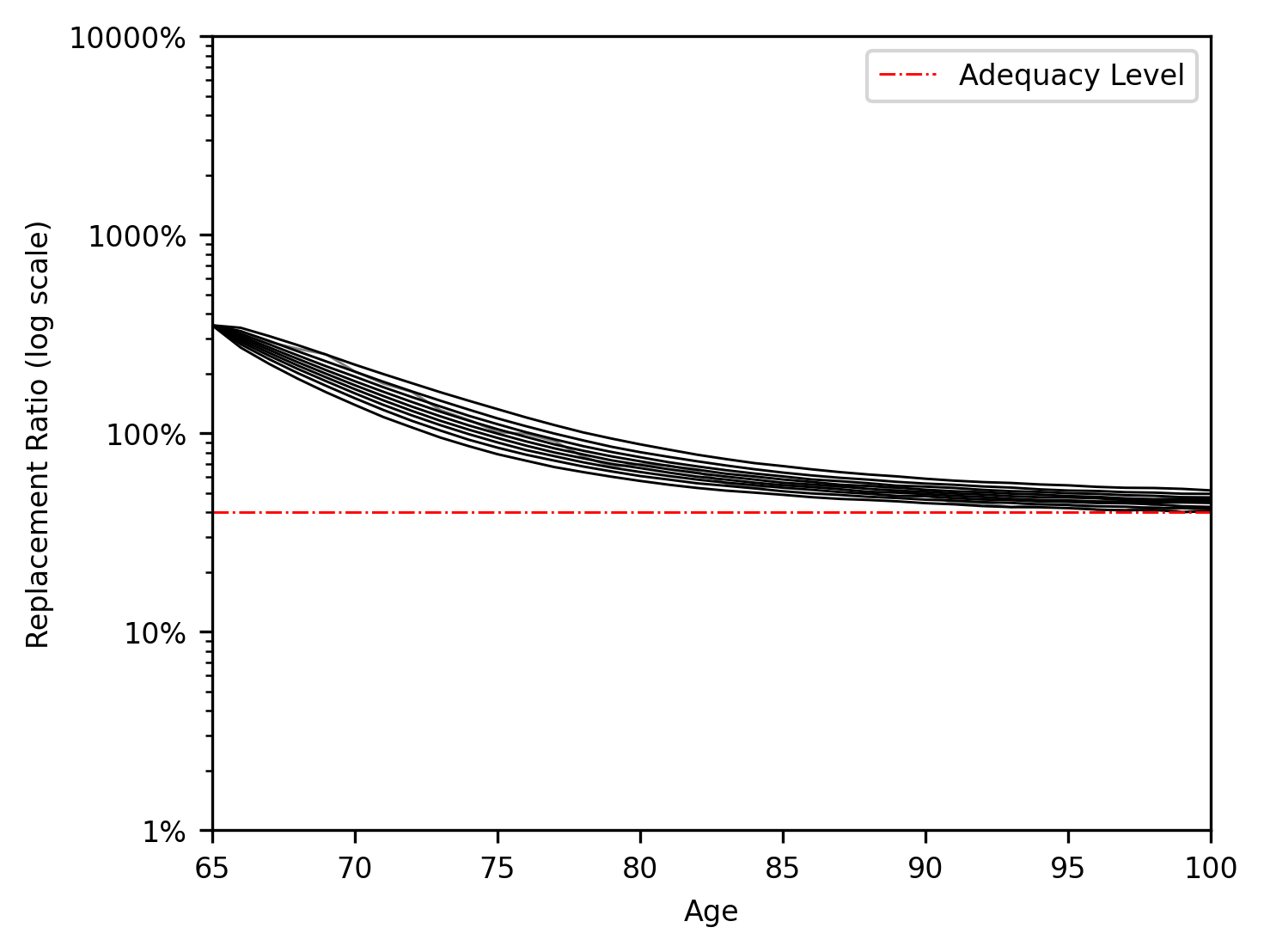}
    \caption{}
    \label{fig:cpp_high_wealth_adequacy}
  \end{subfigure}
  \hfill
  \begin{subfigure}{0.49\textwidth}
    \centering
    \includegraphics[width=\linewidth]{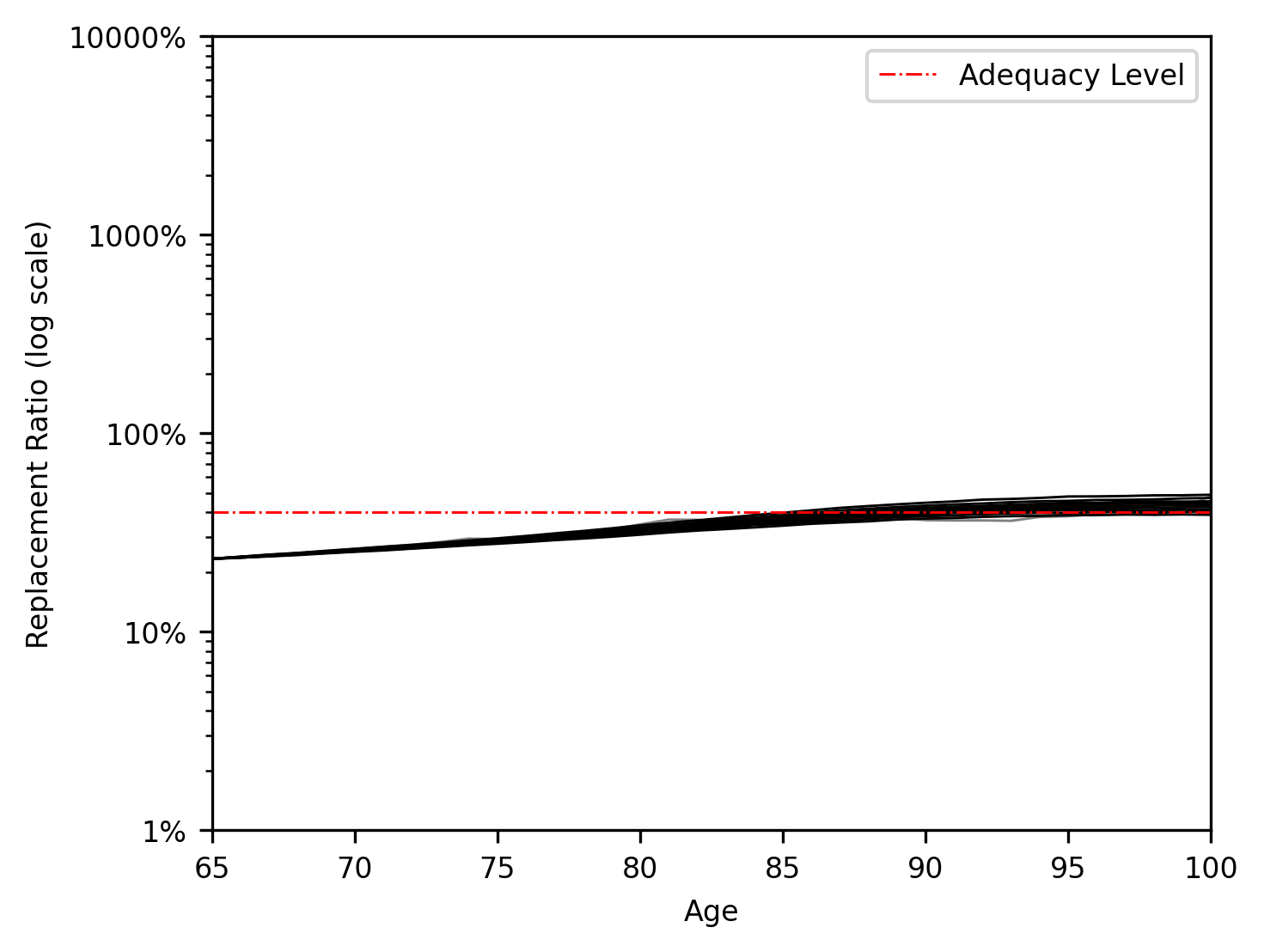}
    \caption{}
    \label{fig:cpp_low_wealth_adequacy}
  \end{subfigure}
  \caption{Comparison of retirement outcomes using the provably convergent numerical method from Appendix \ref{appendix:cvgtMethod}, for highly risk averse individuals in decumulation only. We have Subfigure (\subref{fig:cpp_high_wealth_adequacy}) with high initial wealth and Subfigure (\subref{fig:cpp_low_wealth_adequacy}) with low initial wealth. Each line represents a decile from the simulation.}
  \label{fig:cpp_risk_adequacy_plots}
\end{figure}

This explains the importance of adequacy in the decumulation phase. If one considers accumulation, as we have done throughout, the impact of the adequacy parameter is harder to detect. A highly risk averse individual will invest almost entirely in the risk-free asset during accumulation. This results in an essentially deterministic level of income at retirement, and their consumption throughout retirement will again be determined by whether this is greater or less than $V^{\text{adequate}}$.

\begin{figure}[htbp]
  \centering
  \begin{subfigure}[b]{0.49\textwidth}
    \includegraphics[width=\linewidth]{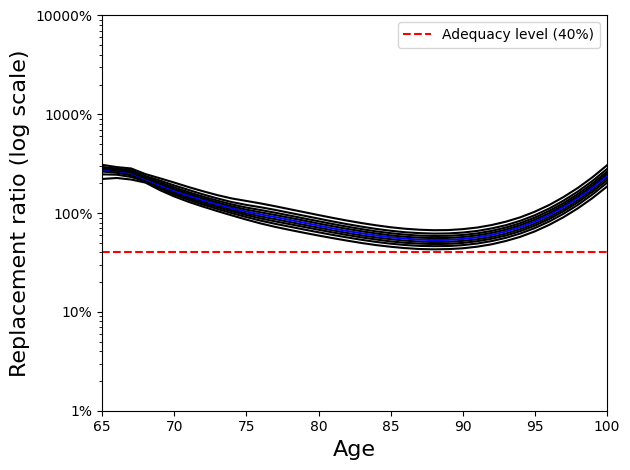}
    \caption{}
    \label{fig:NN_high_wealth_adequacy}
  \end{subfigure}
  \hfill
  \begin{subfigure}{0.49\textwidth}
    \centering
    \includegraphics[width=\linewidth]{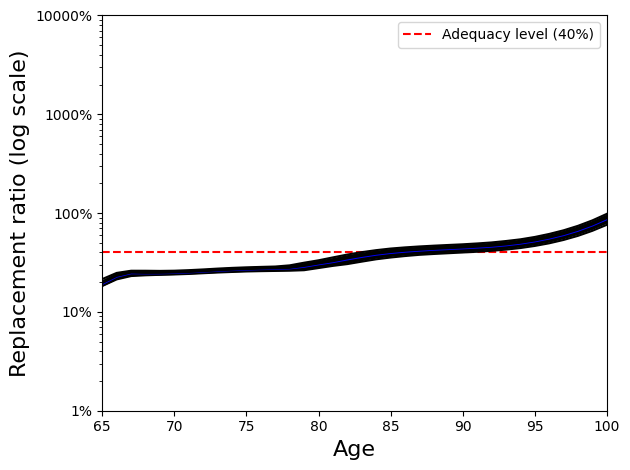}
    \caption{}
    \label{fig:NN_low_wealth_adequacy}
  \end{subfigure}
  
  \vspace{1em}

  \begin{subfigure}[b]{0.49\textwidth}
    \includegraphics[width=\textwidth]{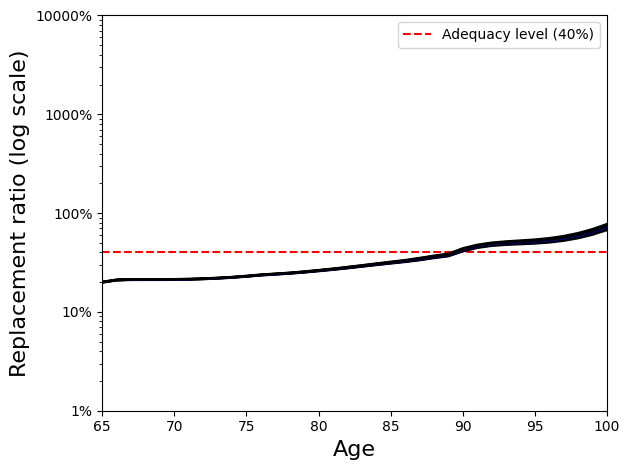}
    \caption{}
    \label{fig:NN_accumulation_adequacy}
  \end{subfigure}
  \caption{Comparison of retirement outcomes when using the neural network for highly risk averse individuals. For the decumulation only setting, we have Subfigure (\subref{fig:NN_high_wealth_adequacy}) with high initial wealth and Subfigure (\subref{fig:NN_low_wealth_adequacy}) with low initial wealth. For an accumulation problem, we have Subfigure (\subref{fig:NN_accumulation_adequacy}). Each line represents a decile from the simulation.}
  \label{fig:NN_risk_adequacy_plots}
\end{figure}

When we view the optimal strategies computed using machine learning, this pattern is somewhat obscured by the fact it is very difficult to train the network to find the optimal strategy beyond the age of about 100 as consumption after this age has only a negligible effect upon utility. The optimal strategies computed using machine learning for decumulation-only are shown in panels (\subref{fig:NN_high_wealth_adequacy}) and (\subref{fig:NN_low_wealth_adequacy}) in Figure \ref{fig:NN_risk_adequacy_plots} and the corresponding accumulation problem in panel (\subref{fig:NN_accumulation_adequacy}). Notice that both decumulation-only strategies differ from the optimum shown in Figure \ref{fig:cpp_risk_adequacy_plots}. The utility values of the low initial wealth problem agree to within 1\%, but the machine learning solution to the high initial wealth problem is not as close to the other numerical method. This is due to the fact that we are using extreme parameters and the gain function becomes hard to compute numerically in these regions. In the accumulation problem, an individual with this level of risk aversion never reaches a wealth above $V^{\text{adequate}}$. As a result, we do not observe the pattern of high initial consumption followed by consumption at the adequacy level. If the individual’s contribution rate (or salary) were sufficiently high, wealth would exceed $V^{\text{adequate}}$ and this behaviour would then emerge. Such contribution rates are, however, somewhat unrealistic in practice.  

If we use realistic values for the risk-aversion parameter, the impact of the adequacy parameter becomes much harder to see. When realistic parameter values are chosen, adjusting the parameter has little visual impact on the optimal strategy. Both the choice of risk-aversion and the choice of adequacy break the positive homogeneity of the problem and it seems that one can use either varying adequacy levels or exponential risk-aversion to avoid the unreasonable strategies found in \cite{armstrong_buescu_dalby} using homogeneous Epstein--Zin preferences.

\section{Conclusions}

In this paper, we have created a tool that allows pension products to be designed by choosing a family of gain functions and then tuning the parameters interactively.

Our main contribution allows us to do this: we control the high variance in gradients that an exponential utility function exhibits across parameter values by scaling the loss. We prove convergence under this scaling for a generic stochastic gradient descent algorithm. We consider two choices of scaling factor, both of which satisfy the assumptions of our convergence result. The first is derived from the convergence bound itself and is therefore theoretically justified but is computationally expensive. The second is a heuristic with no formal justification other than industry standards, but is cheaper and performs well in practice. We compare our methods to a provably convergent algorithm, shown in Appendix \ref{appendix:cvgtMethod}, and confirm their success. Moreover, given the generality of our SGD proof, we expect that our architecture and algorithm could be re-used for more general loss functions and richer economic models across a range of stochastic control problems.

Finally, our tool allows us to analyse economic behaviour of individuals in our setup, and our results are consistent with the literature. We show that the risk aversion and satiation parameters act as expected, but distinguishing between them in the whole of life problem is difficult. The adequacy parameter is more difficult to assess in the whole of life problem, but as we showed in the decumulation-only problem, it is closely tied to the risk aversion of the individual and can restrict consumption in retirement in these settings. We chose the preference model because of Bommier \cite{bommier2006}, and the economic findings from our numerical analysis show that it is consistent with reasonable outcomes.

%% file: pension-appendix.tex
\section{Proof of Theorem \ref{thm:SGD_with_scaling}.}
\label{appendix_scaled_sgd_proof}

Let $x^* \in \argmin L$. By equation \eqref{eq:scaled_expected_utility}, $x^* \in \argmin L_\lambda$ and $\sigma^*_{L_\lambda} = \mathbb{V}\left[\nabla L_{i, \lambda}(x^*) \right]$. For notation, we drop the random arguments of the loss function $L$ such that we write $L(\omega^M, \omega^A, x) = L(x)$.
 
We first establish a one-step contraction valid at an arbitrary fixed scale $\lambda \in [\lambda_\text{min}, \lambda_\text{max}]$, and only afterwards move at each step $k$ to the active scale $\lambda^k$. Expanding the square of the SGD update, we obtain
\begin{equation*}
    \left| \left| x^{k+1}_\lambda - x^* \right| \right|^2 = \left|\left| x^k_\lambda - x^* \right| \right|^2 - 2\gamma \langle \nabla L_{i, \lambda}(x^k_\lambda), x^k_\lambda - x^* \rangle + \gamma^2 \left|\left| \nabla L_{i, \lambda}(x^k_\lambda) \right|\right|^2.
\end{equation*}
By taking conditional expectations, conditioned on $\mathcal{F}^k$, we can obtain
\begin{equation}
    \mathbb{E}_k \left[\left|\left| x^{k+1}_\lambda - x^* \right|\right|^2 \right] = \left|\left| x^k_\lambda - x^* \right| \right|^2 + 2 \gamma \langle \nabla L_\lambda(x_\lambda^k), x^* - x^k_\lambda \rangle + \gamma^2 \mathbb{E}_k \left[ \left|\left| \nabla L_{i, \lambda}(x^k_\lambda) \right|\right|^2 \right].
    \label{eq:expanding_norm}
\end{equation}
By the convexity of $L_\lambda$, we can use Lemma 2.8 from \cite{handbookconvergencetheorems} to see that
\begin{equation*}
    \langle \nabla L_\lambda(x^k_\lambda), x^* - x^k_\lambda \rangle \leq \inf_x L_\lambda(x) - L_\lambda(x^k_\lambda).
\end{equation*}
By applying this and a variance transfer lemma (Lemma 4.20 of \cite{handbookconvergencetheorems}), we can re-write the right-hand-side of \eqref{eq:expanding_norm} to get
\begin{align*}
    \mathbb{E}_k \left[\left|\left| x^{k+1}_\lambda - x^* \right|\right|^2 \right] &\leq \left|\left| x^k_\lambda - x^* \right| \right|^2 + 2\gamma \left(2\gamma \frac{\ell_\text{max}}{\lambda_\text{min}} - 1\right)\left(L_\lambda(x^k_\lambda) - \inf_x L_\lambda(x)\right) + 2 \gamma^2 \sigma^*_{L_\lambda} \\
    & \leq \left|\left| x^k_\lambda - x^* \right| \right|^2 - \gamma\left(L_\lambda(x^k_\lambda) - \inf_x L_\lambda(x)\right) + 2 \gamma^2 \sigma^*_{L_\lambda},
\end{align*}
where in the last inequality we used that $\gamma \leq \frac{\lambda_\text{min}}{4\ell_\text{max}}$. Since $\mathbb{E}\left[\nabla L_{i, \lambda}(x^*)\right] = 0$, the gradient variance is equal to its second moment, and by applying the tower property conditional on $\mathcal{F}^A$ followed by the bound $\lambda(\omega^A) \geq \lambda_\text{min}$ we get
\begin{align*}
    \sigma^*_{L_\lambda} &= \mathbb{E}\left[\left|\left| \nabla L_{i, \lambda}(x^*) \right|\right|^2 \right] \\
    &= \mathbb{E}\left[\frac{\left|\left| \nabla L_i(x^*) \right|\right|^2}{\lambda(\omega^A)^2} \right] \\
    &= \mathbb{E}_{\omega^A}\left[\frac{1}{\lambda(\omega^A)^2}\, \mathbb{E}\left[\left|\left| \nabla L_i(x^*) \right|\right|^2 \,\middle|\, \mathcal{F}^A \right] \right] \\
    &\leq \frac{1}{\lambda^2_\text{min}}\, \mathbb{E}\left[\left|\left| \nabla L_i(x^*) \right|\right|^2 \right] \\
    &= \frac{\sigma^*_L}{\lambda^2_\text{min}}.
\end{align*}
Substituting this in yields, for every admissible fixed scale $\lambda$,
\begin{equation}
    \mathbb{E}_k \left[\left|\left| x^{k+1}_\lambda - x^* \right|\right|^2 \right] \leq \left|\left| x^k_\lambda - x^* \right| \right|^2 - \gamma\left(L_\lambda(x^k_\lambda) - \inf_x L_\lambda(x)\right) + \frac{2 \gamma^2 \sigma^*_L}{\lambda_\text{min}^2}.
    \label{eq:one_step}
\end{equation}
Every bound leading to \eqref{eq:one_step} involves the scale only through the constant $\lambda_\text{min}$, so \eqref{eq:one_step} holds uniformly over all admissible scales $\lambda \in [\lambda_\text{min}, \lambda_\text{max}]$. In particular, at iteration $k$ it holds for the active scale $\lambda^k$:
\begin{equation*}
    \mathbb{E}_k \left[\left|\left| x^{k+1}_\lambda - x^* \right|\right|^2 \right] \leq \left|\left| x^k_\lambda - x^* \right| \right|^2 - \gamma\left(L_{\lambda^k}(x^k_\lambda) - \inf_x L_{\lambda^k}(x)\right) + \frac{2 \gamma^2 \sigma^*_L}{\lambda_\text{min}^2}.
\end{equation*}
Rearranging and taking total expectations gives, for each $k$,
\begin{equation*}
    \gamma\, \mathbb{E}\left[L_{\lambda^k}(x^k_\lambda) - \inf_x L_{\lambda^k}(x)\right]
    \leq \mathbb{E}\left[\left|\left| x^k_\lambda - x^* \right|\right|^2\right]
    - \mathbb{E}\left[\left|\left| x^{k+1}_\lambda - x^* \right|\right|^2\right]
    + \frac{2 \gamma^2 \sigma^*_L}{\lambda_\text{min}^2},
\end{equation*}
in which $\lambda^k$ appears only on the left. Summing over $k = 0, \dots, K-1$ and discarding $-\mathbb{E}\left[\left|\left| x^K_\lambda - x^* \right|\right|^2\right] \leq 0$, we get
\begin{equation}
    \gamma \sum_{k=0}^{K-1} \mathbb{E}\left[L_{\lambda^k}(x^k_\lambda) - \inf_x L_{\lambda^k}(x)\right] \leq \left|\left| x^0_\lambda - x^* \right|\right|^2 + \frac{2 \sigma^*_L}{\lambda_\text{min}^2}\sum_{k=0}^{K-1}\gamma^2.
    \label{eq:telescoped}
\end{equation}
It remains to convert the scaled gaps to the true objective. Since $\inf_x L_{\lambda^k} = L_{\lambda^k}(x^*)$ and $\lambda^k$ is measurable with respect to $\mathcal{F}^k \vee \mathcal{F}^A$, the tower property conditional on $\mathcal{F}^k \vee \mathcal{F}^A$ gives
\begin{equation*}
    \mathbb{E}\left[L_{\lambda^k}(x^k_\lambda) - \inf_x L_{\lambda^k}\right] = \mathbb{E}\left[\frac{1}{\lambda^k(\omega^A)}\, \mathbb{E}\left[L_i(x^k_\lambda) - L_i(x^*) \mid \mathcal{F}^k \vee \mathcal{F}^A\right]\right].
\end{equation*}
By convexity and the fact that $x^*$ is a minimiser, the inner conditional expectation is non-negative, and together with $\lambda^k(\omega^A) \leq \lambda_\text{max}$ this yields
\begin{equation*}
 \frac{1}{\lambda_\text{max}}\, \mathbb{E}\left[L(x^k_\lambda) - \inf_x L\right]
 \leq
     \mathbb{E}\left[L_{\lambda^k}(x^k_\lambda) - \inf_x L_{\lambda^k}\right].
\end{equation*}
Dividing \eqref{eq:telescoped} by $K\gamma$ and applying Jensen's inequality to the fixed convex objective $L$ gives
\begin{equation*}
    \mathbb{E}\left[L(\bar{x}^K_\lambda) - \inf_x L\right] \leq \lambda_\text{max}\left(\frac{\left|\left|x^0_\lambda - x^* \right| \right|^2}{K \gamma} + \frac{2\gamma \sigma^*_L}{\lambda^2_\text{min}} \right),
\end{equation*}
where $\bar{x}_\lambda^K := \frac{1}{K} \sum_{k=0}^{K-1}x^k_\lambda$.
\qed

\section{Proof of Proposition \ref{prop:cvgnce_rate}}
\label{appendix_cvgnce_rate}
For (i), by \eqref{eq:unscaled_conditional_bound}, the right-hand side of the bound is at most $\epsilon(\omega^A)$ for every parameter combination provided
\begin{align*}
    K &\geq \sup_{\omega^A} \left[ \frac{1}{\epsilon^2(\omega^A)} \left( \frac{||x^0-x^*||^2}{\gamma_0} + 2\,\sigma^*_L(\omega^A)\,\gamma_0 \right)^2 \right],
\end{align*}
and since the right-hand side is increasing in the variance and decreasing in the tolerance, \eqref{eq:unscaled_K} suffices. Evaluating this at $\gamma_0 = ||x^0-x^*||/\sqrt{2\sigma_{\max}}$ gives \eqref{eq:unscaled_worst_case}.

For (ii), applying the fixed-scale bound in \eqref{eq:fixed_scale_bound} conditionally on the parameters, with $\sigma^*_{L_\lambda}(\omega^A) = \sigma^*_L(\omega^A)/\lambda^2(\omega^A)$, and noting that conditionally the scale is a constant, so that the left-hand side equals $\lambda(\omega^A)^{-1}\,\mathbb{E}[L(\bar{x}^K_\lambda) - \inf_x L \mid \mathcal{F}^A]$, we obtain \eqref{eq:scaled_conditional_bound}. The bracket in \eqref{eq:scaled_conditional_bound} has the form $A\lambda + B/\lambda$ with $A = ||x^0-x^*||^2/\gamma_0$ and $B = 2\,\sigma^*_L(\omega^A)\,\gamma_0$, and is minimised at $\lambda = \sqrt{B/A}$, which gives \eqref{eq:optimal_scale}.

For (iii), under the choice \eqref{eq:optimal_scale} the scaled gradient variance is the same for every parameter combination, so the single step size $\gamma_0$ is simultaneously optimal for all of them, and the result follows by mirroring Corollary 5.6 of \cite{handbookconvergencetheorems}.
 
\section{Proof of Theorem \ref{thm:scaling_convergence}} \label{appendix_scaling_convergence}
\begin{lemma}[Quadratic bias bound and the cost of depth in moments of $Y$]
\label{rem:quadratic_bias_origin}
Assumptions 1 and 2 give us that there exists $C_B > 0$ such that the gradient biases 
    $B_j^k := -2(T_j^k - T_j^*)$ satisfy
    \begin{equation}
        \mathbb{E}\left[\left|\left| B_j^k \right|\right|^2\right] 
        \leq C_B\left(\sum_{l=1}^{j-1}e_l^k + E_x^k\right) 
        \label{eq:quadratic_bias}
    \end{equation}
    for every $j$ and $k$.
\end{lemma}
\begin{proof}
At component $1$, $T_1^k - T_1^* = Y_i(x^k) - Y_i(x^*)$, so the 
Lipschitz condition alone gives
\begin{equation*}
    \mathbb{E}\left[\left|\left|B_1^k\right|\right|^2\right] 
    = 4\,\mathbb{E}\left[\left|\left|T_1^k - T_1^*\right|\right|^2\right]
    \leq 4C_Y^2\,E_x^k,
\end{equation*}
which is \eqref{eq:quadratic_bias} at $j = 1$ with an empty upstream 
sum. No moments of $Y$ are needed.

At component $j \geq 2$, set $p := T_{j-1}^k - m_{j-1}^k$ and 
$q := T_{j-1}^* - m_{j-1}^*$, and write $G_l := \left|\left|T_l^k - m_l^k\right|\right| 
+ \left|\left|T_l^* - m_l^*\right|\right|$, so that $G_{j-1} = \left|\left|p\right|\right| 
+ \left|\left|q\right|\right|$. Then
\begin{align*}
    \left|\left|T_j^k - T_j^*\right|\right|
    &= \left|\,\left|\left|p\right|\right| - \left|\left|q\right|\right|\,\right| G_{j-1} \\
    &\leq \left|\left|p - q\right|\right| G_{j-1} \\
    &\leq G_{j-1}\left(\left|\left|T_{j-1}^k - T_{j-1}^*\right|\right| 
       + \left|\left|m_{j-1}^k - m_{j-1}^*\right|\right|\right) \\
    &\leq \prod_{l=1}^{j-1}G_l\,\left|\left|T_1^k - T_1^*\right|\right| 
       + \sum_{l=1}^{j-1}\left(\prod_{i=l}^{j-1}G_i\right)
         \left|\left|m_l^k - m_l^*\right|\right| \\
    &\leq W_j\left(C_Y\left|\left|x^k - x^*\right|\right| 
       + \sum_{l=1}^{j-1}\left|\left|m_l^k - m_l^*\right|\right|\right), 
    \qquad W_j := \prod_{l=1}^{j-1}\max\left(1, G_l\right).
\end{align*}
The fourth line iterates the third down to component $1$, and the last uses 
$\prod_{i=l}^{j-1}G_i \leq W_j$ together with the Lipschitz bound 
$\left|\left|T_1^k - T_1^*\right|\right| \leq C_Y\left|\left|x^k - x^*\right|\right|$.
Squaring, applying $\left(\sum_{r=1}^{j}c_r\right)^2 \leq 
j\sum_{r=1}^{j}c_r^2$ to the $j$ summands in the bracket, and noting 
that the bracket is $\mathcal{F}^k$-measurable, the tower property 
gives
\begin{equation*}
    \mathbb{E}\left[\left|\left|B_j^k\right|\right|^2\right] 
    \leq 4j R_j\left(1 \vee C_Y^2\right)
    \left(\sum_{l=1}^{j-1}e_l^k + E_x^k\right), 
    \qquad 
    R_j := \sup_k\,\operatorname*{ess\,sup}\,\mathbb{E}\left[W_j^2 
    \,\middle|\, \mathcal{F}^k\right].
\end{equation*}
Finiteness of $R_j$ follows since $\mathbb{E}\left[W_j^2 \mid \mathcal{F}^k\right]$ 
depends only on $x^k$ and $m_1^k, \ldots, m_{j-1}^k$, confined to bounded sets by the 
compactness of the admissible set and by the truncations of each estimate respectively.
% Having given the derivation of the bias bound, we now show how this affects the requirements on moments of $Y$. Each component squares the 
% residual of the one before it, so $T_l^*$, and hence $G_l$, has 
% degree $2^{l-1}$ in $Y_i(x^*)$, and
% \begin{equation*}
%     \deg W_j^2 = 2\sum_{l=1}^{j-1}2^{l-1} = 2^j - 2 < 2^J.
% \end{equation*}
% The moments of $Y_i(x^*)$ required are therefore of order strictly 
% below $2^J$, supplied by assumption 3 through Lyapunov's inequality. 
% Moments of $Y_i(x^k)$ are controlled by the same via 
% $\left|\left|Y_i(x^k)\right|\right| \leq \left|\left|Y_i(x^*)
% \right|\right| + C_Y\left|\left|x^k - x^*\right|\right|$. Thus the bias admits the quadratic bound specified in Theorem \ref{thm:scaling_convergence}.

% Finiteness of the 
% gradient variances requires moments of degree $2^j \leq 2^J$, since 
% $\sigma_j^* \leq 4\,\mathbb{E}\left[(T_j^*)^2\right]$. Therefore, this is the binding constraint on the moments of $Y$ and the bias bound comes at no moment 
% cost beyond that already spent on the gradient variance.
\end{proof}

Each objective $L_j(m) = \mathbb{E}\left[(T_j^* - m)^2\right]$ has 
gradient $\nabla L_j(m) = 2(m - m_j^*)$ and is therefore $2$-strongly 
convex and $2$-smooth in the sense of 
\cite{handbookconvergencetheorems}. The stochastic gradient at component 
$j$ decomposes as
\begin{equation*}
    g_j^k = -2(T_j^k - m_j^k) = g_j^{k,*} + B_j^k, \qquad 
    g_j^{k,*} := -2(T_j^* - m_j^k),
\end{equation*}
where $g_j^{k,*}$ is an unbiased stochastic gradient of $L_j$ at 
$m_j^k$.
 
At component $1$, by \eqref{eq:quadratic_bias} with an empty upstream 
sum, $\mathbb{E}\left[\left|\left|B_1^k\right|\right|^2\right] \leq 
C_B E_x^k$. By the non-expansiveness of the projection, we can bound $\left|\left|m_1^{k+1} - m_1^*
\right|\right|^2$ by
\begin{equation*}
    \mathbb{E}_k\left[\left|\left|m_1^{k+1} - m_1^*
    \right|\right|^2\right] 
    \leq\left|\left|m_1^k - m_1^*\right|\right|^2 
    - 2\gamma\,\mathbb{E}_k\left[\langle m_1^k - m_1^*, 
    g_1^k\rangle\right] 
    + \gamma^2\,\mathbb{E}_k\left[\left|\left|g_1^k
    \right|\right|^2\right].
\end{equation*}
By Lemma~2.14 of \cite{handbookconvergencetheorems} with strong 
convexity constant $2$, $x = m_1^k$ and $y = m_1^*$:
\begin{equation*}
     \left|\left|m_1^k - m_1^*\right|\right|^2 
    + \left(L_1(m_1^k) - L_1(m_1^*)\right) \leq \langle m_1^k - m_1^*, \nabla L_1(m_1^k)\rangle.
\end{equation*}
Applying the same lemma with $x = m_1^*$, $y = m_1^k$ and 
$\nabla L_1(m_1^*) = 0$ gives 
$\left|\left|m_1^k - m_1^*\right|\right|^2 \leq 
L_1(m_1^k) - L_1(m_1^*)$. Using this with Cauchy-Schwarz and 
Young's inequality $2ab \leq a^2 + b^2$ with 
$a = \sqrt{\gamma(L_1(m_1^k) - L_1(m_1^*))}$ and 
$b = \sqrt{\gamma\,\mathbb{E}_k\left[\left|\left|B_1^k
\right|\right|^2\right]}$:
\begin{equation*}
    2\gamma\left|\mathbb{E}_k\left[\langle m_1^k - m_1^*, 
    B_1^k\rangle\right]\right| 
    \leq \gamma\left(L_1(m_1^k) - L_1(m_1^*)\right) 
    + \gamma\,\mathbb{E}_k\left[\left|\left|B_1^k
    \right|\right|^2\right].
\end{equation*}
For the gradient squared term, using 
$\left|\left|g_1^k\right|\right|^2 \leq 
2\left|\left|g_1^{k,*}\right|\right|^2 + 
2\left|\left|B_1^k\right|\right|^2$ and the Variance Transfer 
Lemma (Lemma~4.20 of \cite{handbookconvergencetheorems}) applied 
to $g_1^{k,*}$ with $L_{\max} = 2$:
\begin{equation*}
    \gamma^2\,\mathbb{E}_k\left[\left|\left|g_1^k
    \right|\right|^2\right] 
    \leq 16\gamma^2\left(L_1(m_1^k) - L_1(m_1^*)\right) 
    + 4\gamma^2\sigma_1^* 
    + 2\gamma^2\,\mathbb{E}_k\left[\left|\left|B_1^k
    \right|\right|^2\right].
\end{equation*}
The total coefficient of $L_1(m_1^k) - L_1(m_1^*)$ is 
$-2\gamma + \gamma + 16\gamma^2 = \gamma(16\gamma - 1)$, 
which is non-positive since $\gamma \leq \frac{1}{16}$, and hence can 
be dropped. This is the binding constraint on the step size. The 
coefficient of $\mathbb{E}_k\left[\left|\left|B_1^k
\right|\right|^2\right]$ is $\gamma + 2\gamma^2 \leq 2\gamma$. 
Taking full expectations and substituting 
\eqref{eq:quadratic_bias}:
\begin{equation}
    e_1^{k+1} \leq (1 - 2\gamma)\,e_1^k + 4\gamma^2\sigma_1^* 
    + 2C_B\,\gamma\,E_x^k. \label{eq:R1_meta}
\end{equation}
At component $j \geq 2$ the weight $\delta_j := \frac{j-1}{J-1} = 2 - a_j$ 
is strictly positive, and the expansion and the treatment of the 
unbiased cross term are identical. For the bias cross term, applying Cauchy-Schwarz followed by the weighted Young inequality 
$2ab \leq \delta_j a^2 + \frac{1}{\delta_j}b^2$ with 
$a = \sqrt{\gamma}\left|\left|m_j^k - m_j^*\right|\right|$ and 
$b = \sqrt{\gamma}\,\mathbb{E}_k\left[\left|\left|B_j^k
\right|\right|^2\right]^{1/2}$, we find
\begin{equation*}
    2\gamma\left|\mathbb{E}_k\left[\langle m_j^k - m_j^*, 
    B_j^k\rangle\right]\right| 
    \leq \delta_j\gamma\left|\left|m_j^k - m_j^*
    \right|\right|^2 
    + \frac{\gamma}{\delta_j}\,\mathbb{E}_k\left[\left|\left|
    B_j^k\right|\right|^2\right].
\end{equation*}
The first term weakens the contraction from $(1 - 2\gamma)$ to 
$(1 - (2 - \delta_j)\gamma) = (1 - a_j\gamma)$, which remains a 
valid contraction since $a_j \leq 2$ and hence 
$a_j\gamma \leq \frac{1}{8} < 1$. The 
gradient squared term is bounded exactly as at component $1$, with 
total loss gap coefficient $-2\gamma + 16\gamma^2 = 
2\gamma(8\gamma - 1) \leq 0$, dropped. The coefficient of 
$\mathbb{E}_k\left[\left|\left|B_j^k\right|\right|^2\right]$ is
\begin{align*}
    \frac{\gamma}{\delta_j} + 2\gamma^2 
    &= \gamma\left(\frac{J-1}{j-1} + 2\gamma\right) \\ 
    &\leq \gamma\left(J - 1 + \tfrac{1}{8}\right)\\
    &\leq J\gamma.
\end{align*}
Taking full expectations and substituting 
\eqref{eq:quadratic_bias}:
\begin{equation}
    e_j^{k+1} \leq (1 - a_j\gamma)\,e_j^k + 4\gamma^2\sigma_j^* 
    + JC_B\,\gamma\left(\sum_{l=1}^{j-1}e_l^k + E_x^k\right). 
    \label{eq:Rj_meta}
\end{equation}
 
We now show by induction on $j$ that there exist constants 
$P_j = \Pi^{j-1}$, $Q_j \leq 4\,\Pi^{j-1}$ and 
$K_j \leq JC_B\,\Pi^{j-1}$ such that for all $k \geq 0$:
\begin{equation}
    e_j^k \leq P_j\,(1 - a_j\gamma)^k\,\Theta_j^0 
    + \gamma\,Q_j\,\Sigma_j^* + \gamma\,K_j\,S_{x,(j)}^k. 
    \label{eq:Hj_meta}
\end{equation}
For the base case, iterating \eqref{eq:R1_meta} from $s = 0$ to 
$s = k-1$ and using 
$\sum_{s=0}^{k-1}(1-2\gamma)^{k-1-s} \leq \frac{1}{2\gamma}$:
\begin{equation*}
    e_1^k \leq (1-2\gamma)^k\left|\left|m_1^0 - m_1^*
    \right|\right|^2 + 2\gamma\sigma_1^* 
    + 2C_B\,\gamma\,S_{x,(1)}^k,
\end{equation*}
so \eqref{eq:Hj_meta} holds for $j = 1$ with $P_1 = 1$, 
$Q_1 = 2$, $K_1 = 2C_B$, since $a_1 = 2$, $\Theta_1^0 = 
\left|\left|m_1^0 - m_1^*\right|\right|^2$ and 
$\Sigma_1^* = \sigma_1^*$. At $k = 0$ 
the bound reduces to $e_1^0 = \left|\left|m_1^0 - m_1^*
\right|\right|^2$ together with non-negative remainders, so it holds 
there also.
 
For the inductive step, fix $j \geq 2$ and assume 
\eqref{eq:Hj_meta} for all $l < j$ and all $k \geq 0$. Iterating 
\eqref{eq:Rj_meta} from $s = 0$ to $s = k-1$:
\begin{align*}
    e_j^k \leq &(1-a_j\gamma)^k\left|\left|m_j^0 - m_j^*
    \right|\right|^2 
    + 4\gamma^2\sigma_j^*\sum_{s=0}^{k-1}(1-a_j\gamma)^{k-1-s} \\ 
    &+ JC_B\,\gamma\sum_{s=0}^{k-1}(1-a_j\gamma)^{k-1-s}
    \left(\sum_{l<j}e_l^s + E_x^s\right).
\end{align*}
Since $a_j \geq 1$, $\sum_{s=0}^{k-1}(1-a_j\gamma)^{k-1-s} \leq 
\frac{1}{a_j\gamma} \leq \frac{1}{\gamma}$, so the variance term 
is at most $4\gamma\sigma_j^* \leq 4\gamma\Sigma_j^*$; the 
initialisation term is at most $(1-a_j\gamma)^k\Theta_j^0$; and 
the term in $E_x^s$ is by definition $JC_B\,\gamma\,S_{x,(j)}^k$. 
For the upstream terms, substituting \eqref{eq:Hj_meta} for each 
$l < j$ and using the monotonicity $\Theta_l^0 \leq \Theta_j^0$ 
and $\Sigma_l^* \leq \Sigma_j^*$ produces three components, each 
handled via the convolution formula
\begin{equation*}
    \sum_{s=0}^{k-1}r_1^{k-1-s}\,r_2^s 
    = \frac{r_1^k - r_2^k}{r_1 - r_2}
\end{equation*}
with $r_1 = 1 - a_j\gamma$ and $r_2 = 1 - a_l\gamma$, which are 
distinct for every $l < j$ since $a_l > a_j$ strictly, with 
$r_1 - r_2 = (a_l - a_j)\gamma \geq \frac{\gamma}{J-1} > 0$. For the initialisation component:
\begin{align*}
    JC_B\,\gamma\,P_l\,\Theta_j^0\sum_{s=0}^{k-1}
    (1-a_j\gamma)^{k-1-s}(1-a_l\gamma)^s 
    &\leq JC_B\,\gamma\,P_l\,\Theta_j^0\,
    \frac{(1-a_j\gamma)^k}{(a_l - a_j)\gamma} \\ 
    &\leq J(J-1)C_B\,P_l\,(1-a_j\gamma)^k\,\Theta_j^0.
\end{align*}
For the gradient variance component:
\begin{equation*}
    JC_B\,\gamma^2\,Q_l\,\Sigma_j^*
    \sum_{s=0}^{k-1}(1-a_j\gamma)^{k-1-s} 
    \leq JC_B\,\gamma\,Q_l\,\Sigma_j^*.
\end{equation*}
For the moving target component, substituting 
$S_{x,(l)}^s = \sum_{i=0}^{s-1}(1-a_l\gamma)^{s-1-i}E_x^i$, 
exchanging the order of summation, substituting $t = s - i - 1$ 
in the inner sum and applying the convolution formula:
\begin{align*}
    \sum_{s=i+1}^{k-1}(1-a_j\gamma)^{k-1-s}
    (1-a_l\gamma)^{s-1-i} 
    &= \frac{(1-a_j\gamma)^{k-1-i} - (1-a_l\gamma)^{k-1-i}}{(a_l - a_j)\gamma} \\ 
    &\leq \frac{(J-1)(1-a_j\gamma)^{k-1-i}}{\gamma},
\end{align*}
so this component is bounded by 
$J(J-1)C_B\,\gamma\,K_l\,S_{x,(j)}^k$. Collecting the direct and 
upstream contributions, \eqref{eq:Hj_meta} holds with:
\begin{equation*}
    P_j = 1 + J(J-1)C_B\sum_{l<j}P_l, \qquad 
    Q_j = 4 + JC_B\sum_{l<j}Q_l, \qquad 
    K_j = JC_B + J(J-1)C_B\sum_{l<j}K_l.
\end{equation*}
Writing $\Lambda_j := \sum_{l<j}P_l$ gives $\Lambda_1 = 0$ and 
$\Lambda_{j+1} = \Pi\,\Lambda_j + 1$, which solves to 
$\Lambda_j = \frac{\Pi^{j-1}-1}{\Pi - 1}$ and hence 
$P_j = 1 + (\Pi - 1)\Lambda_j = \Pi^{j-1}$. By induction using 
$JC_B \leq \Pi - 1$:
\begin{equation*}
    Q_j \leq 4 + 4JC_B\,\frac{\Pi^{j-1}-1}{\Pi-1} 
    \leq 4\,\Pi^{j-1},
\end{equation*}
and identically $K_j \leq JC_B\,\Pi^{j-1}$. Putting 
everything together gives the bound in the statement of the theorem.
\qed

\section{Log Transformations of Loss Functions}
\label{appendix_log_transformations}
To mitigate the numerical instability induced by the exponential utility function in our problem formulation, we apply logarithmic transformations throughout the implementation. Although motivated by the specific structure of our utility function, these transformations are applicable more generally to problems that suffer from numerical overflow in their implementation.

Whenever possible, quantities are maintained in log space to avoid overflow and improve numerical stability. Loss evaluations are then performed using numerically stable operations such as \textit{logsumexp}. For squared differences, we use the identity
\begin{equation*}
\log\left((e^a-e^b)^2 \right)
= 2 \left[\, \max(a, b)
+ \log1\text{p}\bigl(- e^{\min(a, b) - \max(a, b)}\bigr) \right],
\label{eq:logsquareddiffexp}
\end{equation*}
where $\log1\text{p}(x):=\log(1+x)$.

Applying these transformations yields the following loss function for the main RNN:
\begin{multline}\label{eq:log_transform_loss}
    L = \log\left(\sum^N_{s=1}\exp\left(\log\left(\sum_{t=t_{\text{RA}}}^T \exp\left(\log(\overline{p}_t)-\alpha \sum^t_{j=t_{\text{RA}}}u(C_{j}^{(s)}) \delta t\right)\right) - \log\left( \lambda^k_s \right)\right) \right) \\
    - \log(N),
\end{multline}
where $\lambda^k_s$ is the current (k-th) estimate of the scale on the $s$-sample parameters.

We also apply logarithmic transformations to the auxiliary network loss functions when appropriate. For the gradient-based method, the loss $L_\mu$ is used in the form given in Equation \eqref{eq:L_mu}. In contrast, for the heuristic method we use a log-transformed version because the loss magnitude can become extremely large:
\begin{equation*}
    L_\mu(\omega^A) = \log\left(\sum_{s=1}^N \exp\left(\log\left(\left( \exp(\log(Y_s)) - \exp(\log(\mu_\theta(\omega^A_s))) \right)^2\right)\right) \right) - \log(N),
\end{equation*}
where $Y_s$ is the specific variable we are trying to estimate the mean and variance of.

We do not apply a logarithmic transformation in the gradient-based setting because the gradients themselves do not exhibit the same degree of numerical instability as the loss values. Moreover, gradients may take negative values and therefore cannot, in general, be used directly as arguments of the logarithm. By contrast, the heuristic loss is strictly positive as it is an exponential, allowing the transformation to be applied safely.

For both methods, the variance-related loss $L_v$ can be log-transformed for improved numerical stability, since both the variance and squared residual terms are strictly positive. This gives
\begin{equation*}
    L_v(\omega^A) = \log\left(\sum_{s=1}^N \exp\left(\log\left(\left( \exp(\log(R_s)) - \exp(\log(v_\theta(\omega^A_s))) \right)^2\right)\right)\right) - \log(N).
\end{equation*}

\section{Utility Function Parameter Ranges}
\label{appendix_table_params}

\begin{table}[htbp]
\centering
\begin{tabular}{c|c}
\toprule
\textbf{Parameter} & \textbf{Value} \\
\midrule
$\alpha$ & $10^{-5}$ \\
$\rho$   & $-2.0$ \\
$a$      & $0.4$ \\
\bottomrule
\end{tabular}
\caption{Default Parameter Selection}
\label{tab:parameters}
\end{table}

\begin{table}[htbp]
\centering
\begin{tabular}{c|c}
\toprule
\textbf{Parameter} & \textbf{Range} \\
\midrule
$\alpha$ & $[10^{-7}, 0.2]$ \\
$\rho$   & $[-2.0, -0.01]$ \\
$a$      & $[0.01, 1.0]$ \\
\bottomrule
\end{tabular}
\caption{Parameter ranges}
\label{tab:param_ranges}
\end{table}

\section{Neural Network Architecture Details}
\label{appendix_NN}
Our code is written in Python using the Tensorflow package. 
\subsection{RNN Architecture}
The main RNNs used in the fixed-parameter method, and both the gradient-based and heuristic methods are identical aside from the inputs. This architecture has not been heavily optimised and is knowingly over engineered in that it does not exploit the Markovianity of the problem. We simply want to produce a tool that can be used when the problem is non-Markovian, and so provides a check that Markovianity is not essential.

We mark in brackets the additional inputs to the varying-parameter networks. The networks consist of six layers: 
\begin{itemize}
    \item The first input layer has two (five) nodes, representing: the Gaussian increments and the time points (and the three parameter values).

    \item The second layer is a dense layer with $80$ nodes and a ReLU activation function.

    \item The third layer is a gated recurrent unit (GRU) with $25$ nodes. This is the recurrent layer in our network. The activation function is the hyperbolic tangent function and the recurrent activation function is sigmoid. The GRU ensures that the network returns an output at each time point.
    
    \item The fourth and fifth layers are identical to the second layer.

    \item The final output layer has two nodes, which represent the proportion of wealth to invest in the risky asset and the proportion of wealth to consume. We use the linear activation function for both outputs, and perform a transformation of the consumption proportion so that it remains in the interval $[0, 1]$. We obtain investment and consumption decisions in each year of our simulation as a result of the GRU.

\end{itemize}
\begin{figure}[htp!]
  \centering
  \begin{minipage}[b]{0.80\textwidth}
    \centering
    \includegraphics[width=1.0\linewidth]{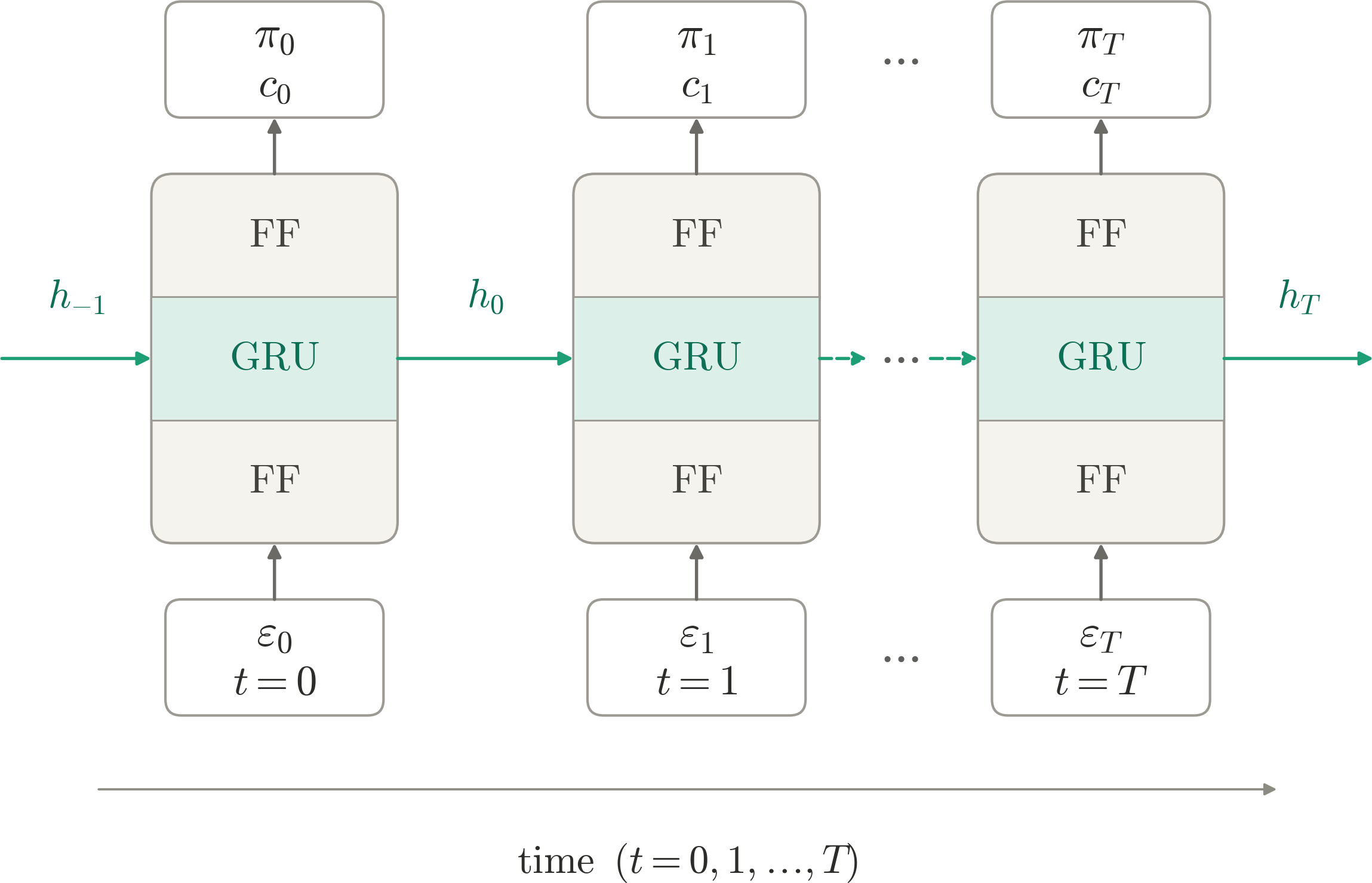}
  \end{minipage}
    \caption{Architecture for the Recurrent Neural Network. All the layers are dense. The label `FF' denotes a feedforward layer.}
    \label{fig:RNN_architecture}    
\end{figure}

\noindent
We used the Adam optimizer with an initial learning rate of $0.001$. In every case, training was carried out for 24 hours, each epoch consisting of $131,072$ scenarios with a batch size of $4,096$. In the gradient-based method, computational capacity meant that only batch sizes of size $128$ could be used. A validation set of $10,000$ separately generated scenarios was evaluated at the end of each epoch for the non gradient-based methods. We allow all the networks to train for a large number of epochs and simply extract the weights for which the validation loss was least.

\subsection{The Auxiliary Networks}
The `scaling' network in both methods is a feedforward neural network with a much simpler architecture. The mean estimating network in the heuristic method has the same simple architecture, but in the gradient-based method, the output layer must match the number of entries of the gradient vector, i.e. the number of parameters of the main RNN. All the networks consist of an input layer of three nodes that takes the three parameter values. They all have three hidden layers each with $64$ nodes. All of the hidden layers use the ReLU activation function. In the case of the scaling networks, and the mean predicting network in the heuristic method, the output layer has one node, representing predictions for the variance or the mean of the desired variable. The mean network in the gradient-based method requires an output layer equal to the number of parameters of the main neural network, since it attempts to learn the average value of each of the entries of the gradient vector conditional on a parameter combination. The output layer uses the linear activation function since we take the scaling factors as logarithms.

All of these networks are trained in the same loop as the main RNN, and so have the same training parameters as the main RNN.

\subsection{Replacement Ratio Percentile Network}
The replacement ratio percentile network is a feedforward neural network. It consists of an input layer with five nodes, to represent the time point, the percentile and the three varying parameters. So the replacement ratio percentile network learns the value of the replacement ratio for a given time point, in a given percentile, for a given set of parameters. There are three hidden layers consisting of 64 nodes, all with the ReLU activation function. The output layer has a single node as the network only makes one prediction per data point. The output layer uses the sigmoid activation function since we transform both inputs and outputs to the interval $[0, 1]$. Each data point is one specific time point from one of the deciles. We only need $50$ epochs to find the minimum. 

\section{Neural Network Comparison \& Strategy Plots}
\label{AppendixPlots}
Here, we first show comparisons of the outcomes for a fixed set of parameters for the two respective methods in the paper as compared to fixed networks. We then also show the strategies associated with the sensitivity analysis plots in Section \ref{sec:sensitivity_analysis}.

\begin{figure}[htp!]
  \centering
  \begin{subfigure}[b]{0.48\textwidth}
    \centering
    \includegraphics[width=\linewidth]{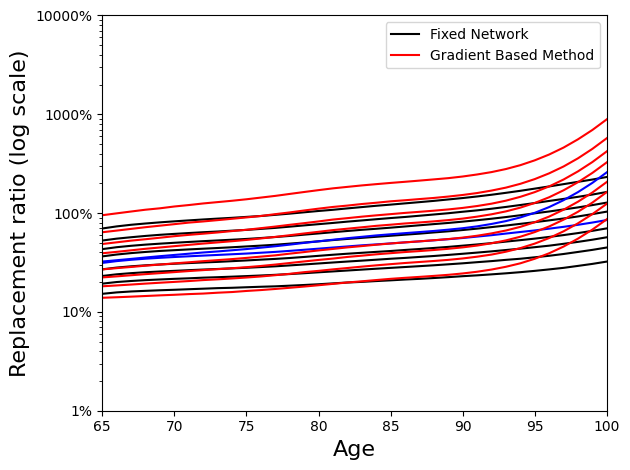}
    \caption{}
    \label{fig:gradient_based_vs_fixed_rr}
  \end{subfigure}
  \hfill
  \begin{subfigure}[b]{0.48\textwidth}
    \centering
    \includegraphics[width=\linewidth]{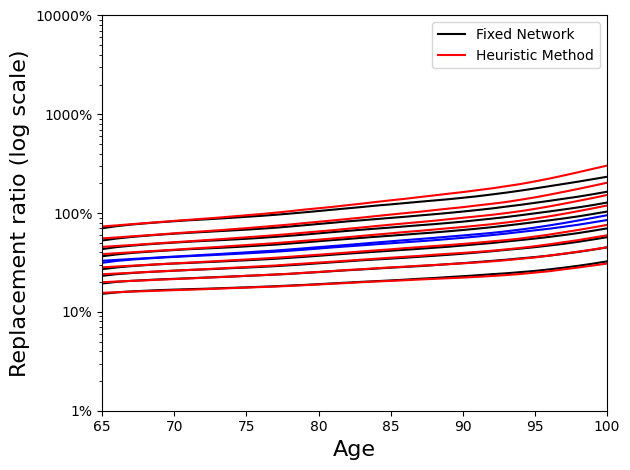}
    \caption{}
    \label{fig:heuristic_vs_fixed_rr}
  \end{subfigure}
  \caption{Panel (\subref{fig:gradient_based_vs_fixed_rr}) shows a comparison of the replacement ratio deciles between the gradient-based method and a fixed parameter network for the default parameters in Table \ref{tab:parameters}. Panel (\subref{fig:heuristic_vs_fixed_rr}) shows the corresponding plot for the heuristic method.}
  \label{fig:comparisons_vs_fixed_rrs}
\end{figure}

\begin{figure}[htp!]
  \centering
  \begin{subfigure}[b]{0.36\textwidth}
    \centering
    \includegraphics[width=\linewidth]{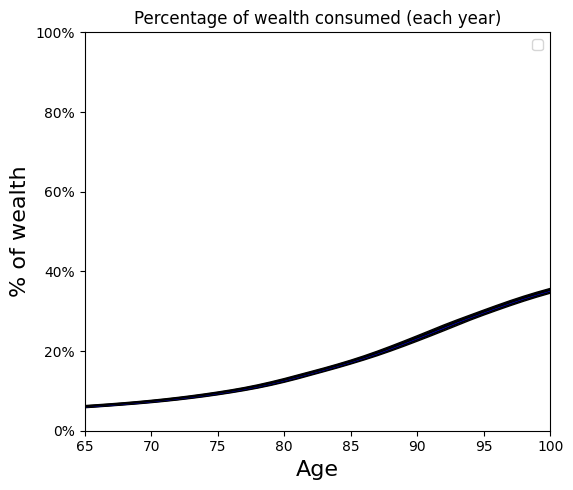}
    \caption{}
    \label{fig:high_alpha_consumption_strategy}
  \end{subfigure}
  \hfill
  \begin{subfigure}[b]{0.62\textwidth}
    \centering
    \includegraphics[width=\linewidth]{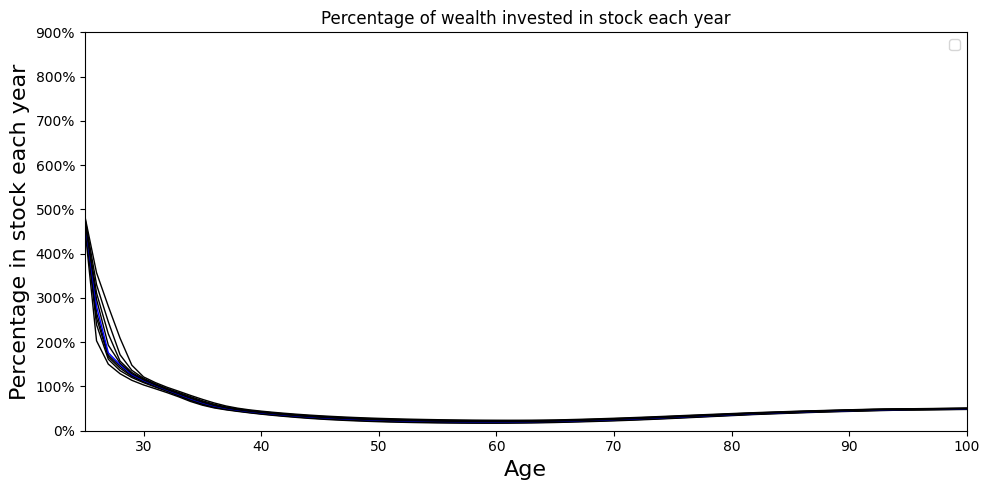}
    \caption{}
    \label{fig:high_alpha_investment_strategy}
  \end{subfigure}
  \caption{Panel (\subref{fig:high_alpha_consumption_strategy}) shows the consumption strategy for the outcomes plotted in Figure \ref{fig:high_alpha_plot}, for high risk aversion. Panel (\subref{fig:high_alpha_investment_strategy}) shows the corresponding investment strategy.}
  \label{fig:high_alpha_strategyies}
\end{figure}

\begin{figure}[htp!]
  \centering
  \begin{subfigure}[b]{0.36\textwidth}
    \centering
    \includegraphics[width=\linewidth]{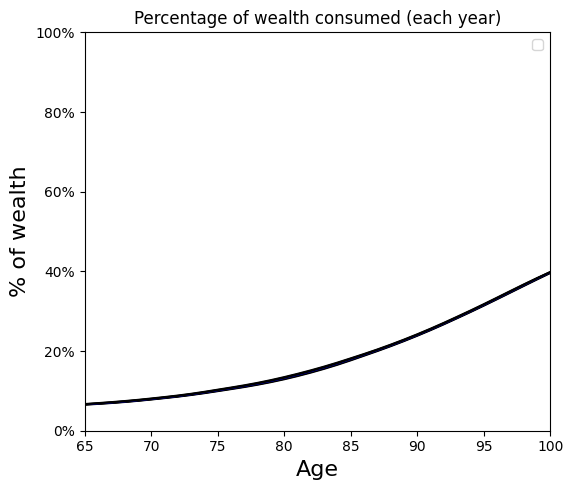}
    \caption{}
    \label{fig:low_alpha_consumption_strategy}
  \end{subfigure}
  \hfill
  \begin{subfigure}[b]{0.62\textwidth}
    \centering
    \includegraphics[width=\linewidth]{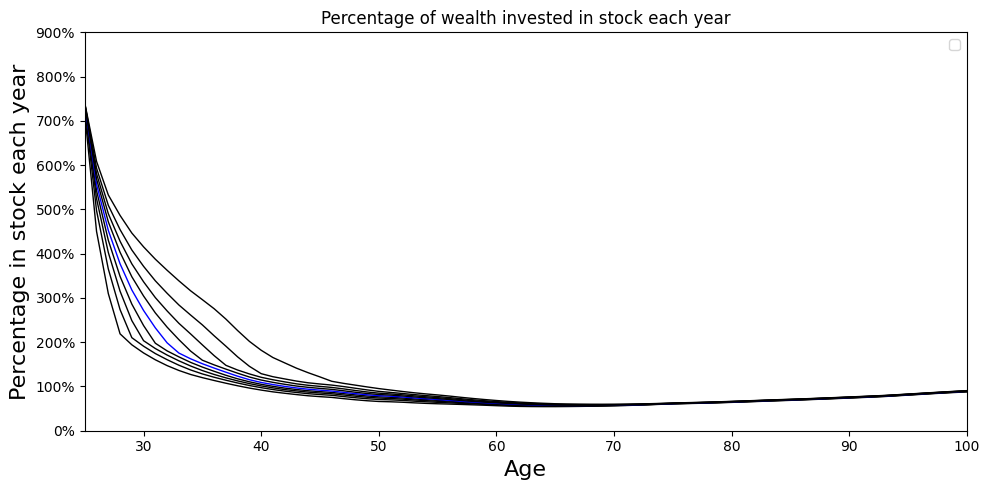}
    \caption{}
    \label{fig:low_alpha_investment_strategy}
  \end{subfigure}
  \caption{Panel (\subref{fig:low_alpha_consumption_strategy}) shows the consumption strategy for the outcomes plotted in Figure \ref{fig:low_alpha_plot}, for low risk aversion. Panel (\subref{fig:low_alpha_investment_strategy}) shows the corresponding investment strategy.}
  \label{fig:low_alpha_strategyies}
\end{figure}

\begin{figure}[htp!]
  \centering
  \begin{subfigure}[b]{0.36\textwidth}
    \centering
    \includegraphics[width=\linewidth]{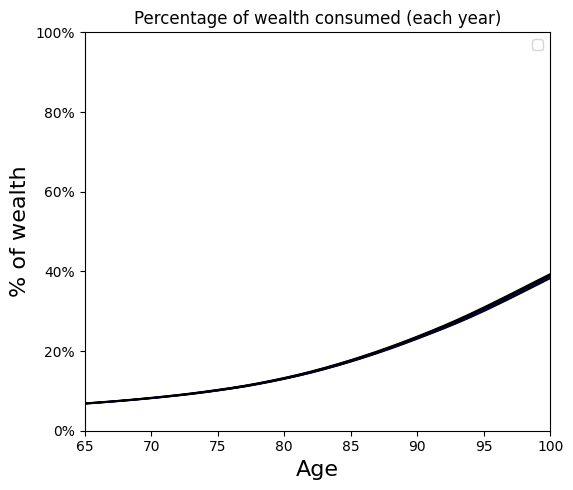}
    \caption{}
    \label{fig:high_rho_consumption_strategy}
  \end{subfigure}
  \hfill
  \begin{subfigure}[b]{0.62\textwidth}
    \centering
    \includegraphics[width=\linewidth]{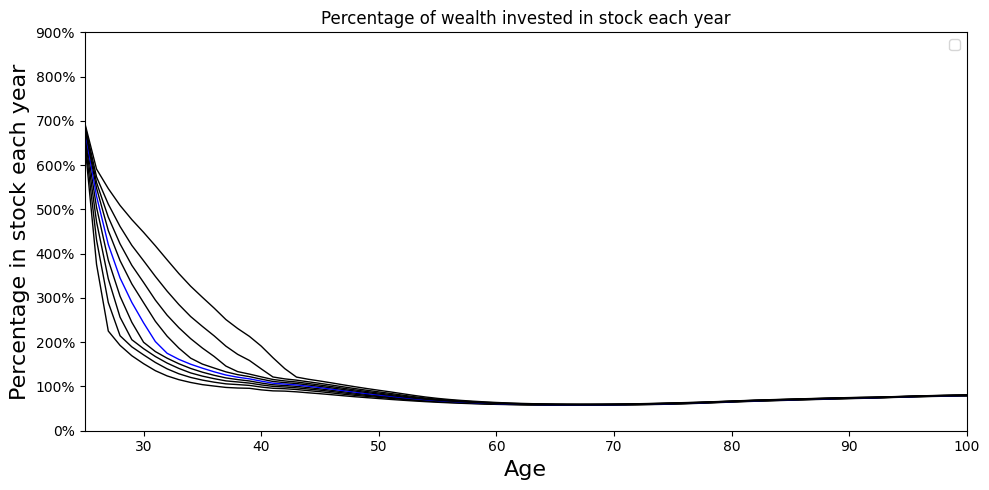}
    \caption{}
    \label{fig:high_rho_investment_strategy}
  \end{subfigure}
  \caption{Panel (\subref{fig:high_rho_consumption_strategy}) shows the consumption strategy for the outcomes plotted in Figure \ref{fig:high_rho_plot}, for more easily satiated preferences. Panel (\subref{fig:high_rho_investment_strategy}) shows the corresponding investment strategy.}
  \label{fig:high_rho_strategyies}
\end{figure}

\begin{figure}[htp!]
  \centering
  \begin{subfigure}[b]{0.36\textwidth}
    \centering
    \includegraphics[width=\linewidth]{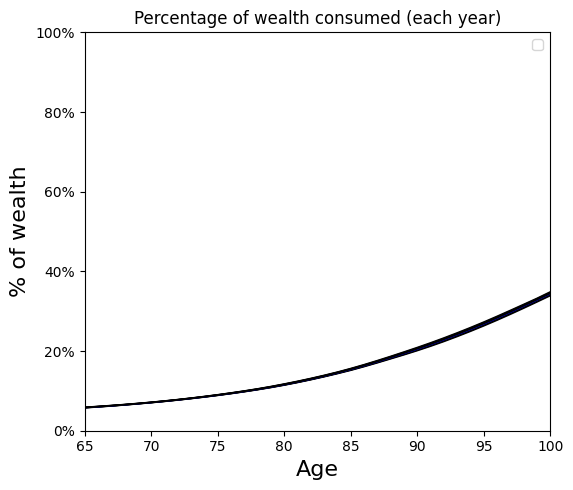}
    \caption{}
    \label{fig:low_rho_consumption_strategy}
  \end{subfigure}
  \hfill
  \begin{subfigure}[b]{0.62\textwidth}
    \centering
    \includegraphics[width=\linewidth]{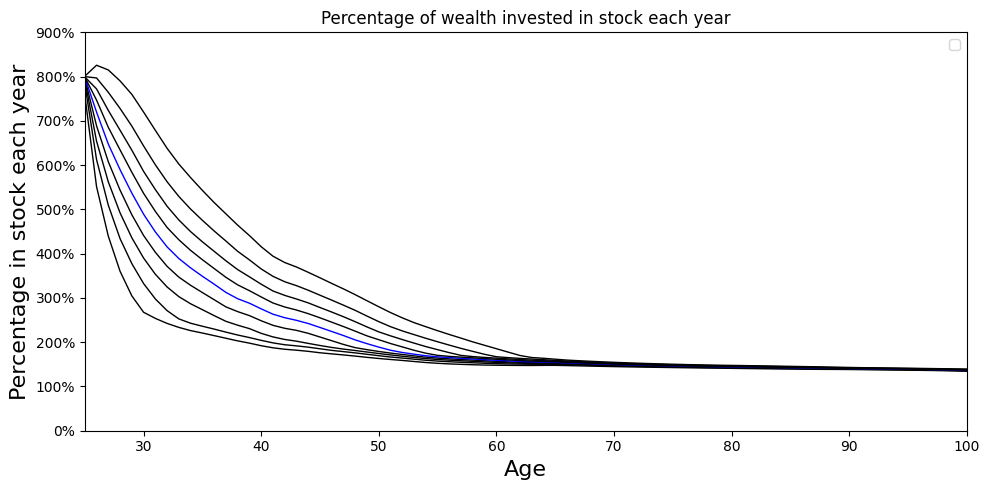}
    \caption{}
    \label{fig:low_rho_investment_strategy}
  \end{subfigure}
  \caption{Panel (\subref{fig:low_rho_consumption_strategy}) shows the consumption strategy for the outcomes plotted in Figure \ref{fig:low_rho_plot}, for less easily satiated preferences. Panel (\subref{fig:low_rho_investment_strategy}) shows the corresponding investment strategy.}
  \label{fig:low_rho_strategyies}
\end{figure}

%% file: classical-method.tex
\section{A Convergent Algorithm for the Discrete Consumption, Continuous Investment Problem in Decumulation}
\label{sec:numericalMethod}
\label{appendix:cvgtMethod}

Recall that the dynamics of $w$ are determined by equations \eqref{eq:returns} and \eqref{eq:wealthInfinite}.
We define an admissible control
to be a progressively measurable process $((C_t)_{t\in{\cal T}},(\pi_t)_{t \in [0,T]})$ such that $w_{t-} \geq 0$
and $w_t \geq 0$  for all time.
We write ${\cal A}$ for the set of admissible controls.

Our objective is to compute
\begin{equation}
v := 
\sup_{(C,\pi) \in {\cal A}} U( C ),
\label{eq:fundObjective}
\end{equation}
and to find $(C, \pi)$ achieving (or if necessary, approximating) this supremum.

Our strategy is to solve the one-period problem using a duality method which will 
allow us to identify the solution with minimal assumptions on the form of our utility
function. To simplify the duality argument, we use the theory of isomorphic markets to recast the problem in a particularly simple form. Having obtained this solution, we
will recursively solve the multi-period problem. 
Our goal in this appendix is to give
all details needed to implement the resulting algorithm and a proof of its convergence.

\subsection{Solution to the One-Period Problem}

Write ${\cal A}_{w,t}$ for the admissible consumption-investment
strategies that start with wealth $w$ at time $t$.  
Define the value function $v$, as
a function of initial wealth, $w$ at time  $t_1 \in {\cal T}$ by
\begin{align*}
v_{t}(w)&:=\sup_{{C,\pi} \in {\cal A}_{w,t}}
    \mathbb{E}\left( -\exp\left(-\alpha \sum_{j \in {\cal T}, \, {t \leq j<\tau}} u(C_j) \delta t\right)\right)
\end{align*}
To make the limits in the sum easier to read, we will write the sum
using the following integral notation
\begin{align*}
v_{t}(w)&=\sup_{{C,\pi} \in {\cal A}_{w,t}}
    \mathbb{E}\left( -\exp\left(-\alpha \int_{t}^{\tau} u(C_s) \ed {\cal T} (s) \right) \right).
\end{align*}

Given $v_{t}$, we wish to compute $v_{t-\delta t}$, we will then be able
to recursively compute $v_{t}$ for all $t \in {\cal T}$. Our next proposition
shows how to compute $v_{t-\delta t}$, but
in order to state our results concisely we first make the following definitions.
\begin{definition}
	Let $f:\R \to \R \cup \{ \pm \infty \}$ be concave and increasing. Define
	\[
	f^\dagger:\R_{>0} \to \R
	\]
	by 
	\[
	f^\dagger(p) = \inf \{ x \mid p \in \partial f(x) \}
	\]
	where $\partial f(x)$ is the sub-differential of $f$ at $x$.
	\label{def:dagger}	
\end{definition}

For sufficiently regular
functions $f$, we have $f^\dagger = (f^\prime)^{-1}$, or, equivalently, $f^\dagger$
is the derivative of the Legendre transform of $f$.

\begin{definition}
Define
\begin{equation}
Q(z):=\Phi\left( M  + \Phi^{-1}(z) \right),
\label{eq:defQ}
\end{equation}
where $\Phi$ is the cumulative distribution function of the
standard normal distribution and
\[
M:= \frac{\left| \mu-r \right|\sqrt{\delta t}}{\sigma }.
\]
Define
\begin{equation}
q^A_{\BS}(z)= \frac{\ed Q}{\ed z}.
\end{equation}
\end{definition}

As we will show in Lemma \ref{lemma:bsmCanonicalForm} below, the quantity $q^A_{\BS}(z)$ can be related to the pricing kernel of the Black--Scholes model.

We may now state the following result which allows us to solve the one period problem.

\begin{proposition}
	Suppose that $t_1=t_0+\delta t$ and that  $v(w):=v_{t_1}(w)$ is known, concave and increasing
	for $w>0$, equal to $-\infty$ for $w\leq 0$,
	and satisfies $v(w) \leq 0$. Define $s_t=(1-p_t)$ for $t \in {\cal T}$, so $s_t$ denotes the survival
    probability over the period $[t,t+\delta t)$.
	\begin{enumerate}[label=(\alph*)]
	\item 	\label{prop:solutionExponentialA} $v_{t_0}(w)$ is itself concave and increasing for $w>0$, equal to $-\infty$ for $w \leq 0$
	and satisfies $v_{t_0}(w) \leq 0$.	
	\item \label{prop:solutionExponentialB}
    For each $\eta>0$ define a function 
	$f^\eta:(0,1)\to \R_{\geq0}$ by
	\begin{equation}
	f^\eta(s) = v^\dagger \left( 
	\eta   e^{-r \delta t} q^A_{\BS}(s) \right).
	\label{eq:defFEta}
	\end{equation}
	Define $C^\eta \in \R_{\geq0}$ by
	\begin{equation}
	C^\eta  = u^\dagger \left( -\frac{\eta}{\delta t}
	\left(
	-1
	+ s_{t_0} \int_0^1 (1+v(f^\eta(s))) \, \ed s \right)^{-1}
	\right).
	\label{eq:defCEta}
	\end{equation}
	Define $w^\eta$ by
	\begin{equation}
	w^\eta = C^{\eta} + s_{t_0} \int_0^1 e^{-r \delta t} q^{A}_{\BS}(s) f^{\eta}(s) ds.
	\label{eq:defWeta}
	\end{equation}
	If there exists $\eta_{w_{t_0}}$ such that $w^{\eta_{w_{t_0}}}=w_{t_0}$ then we have
	\[
	v_{t_0}(w_{t_0}) = 
	\exp( -u(C^{\eta_{w_{t_0}}})  \delta t ) \left(
	-1
	+ s_{t_0} \int_0^1 (1+v(f^{\eta_{w_{t_0}}}(s))) \, \ed s \right)
	\]
	and $C^{\eta_{w_{t_0}}}$ is the optimal consumption at time $t_0$.		
	\end{enumerate}
	\label{prop:solutionExponential}
\end{proposition}

Part \ref{prop:solutionExponentialA} is trivial. For example the
statement about concavity follows from \cite{luenberger} Proposition 8.3.1.
The proof strategy for Part \ref{prop:solutionExponentialB} is as follows:
\begin{enumerate}[label=(\roman*)]
	\item Use the dynamic programming principle to obtain a recursive formulation of the problem.  This is done in Lemma \ref{lemma:dpp}
	\item Reduce the continuous time investment problem of the recursion step to a calculus of variations problem using the classification of one-period complete markets. This is done in Lemma \ref{lemma:reduceToCalculusVariations}.
	\item Solve the resulting calculus of variations problem. This is done in Lemma \ref{lemma:solveCalculusOfVariations}.
\end{enumerate}

Let us first see how to compute 
$v_{t_0}(w_{t_0})$ as the solution to a one period optimal investment problem.

\begin{lemma}
	\label{lemma:dpp}
	Assume the conditions of Proposition \ref{prop:solutionExponential}.
	Let ${\cal A}_{w_{t_0},t_0,t_1}$ denote the set of pairs $(C_{t_0},\pi)$
	where $\pi$ is an admissible investment strategy for the period $[t_0,t_1]$ and $C_{t_0}\in \R$ is the consumption at time $t_0$
	and satisfies $C_{t_0}<w_{t_0}$. Then
	\begin{equation}
	\begin{split}
	v_{t_0
	}(w_{t_0}):=\sup_{{C_{t_0},\pi} \in {\cal A}_{w_{t_0},t_0,t_1}}
	\Big\{
	\exp\left(-u(C_{t_0}) \delta t \right)
	\left( -1 + s_{t_0} 
	\E\left( 1+v_{t_1}(w^{(C_{t_0},\pi)}_{t_1}) \right)
	\right)
	\Big\}
	\label{eq:exponentialDynamicProgramming}
	\end{split}
	\end{equation}
	where $w^{(C_{t_0},\pi)}_{t_1}$ is the value obtained by 
	following the investment strategy $\pi$ from $t_0$ to $t_1$
	with an initial wealth of $s_{t_0}^{-1}(w_{t_0} - C_{t_0})$.
\end{lemma}
\begin{proof}
	We calculate
	\begin{align*}
	v_{t_0}(w_{t_0})
	&=\sup_{{C,\pi} \in {\cal A}_{w_{t_0},t_0}} \Big\{ 
	\E\left(-\exp\left(-u(C_{t_0}) \delta t \right) \P(\tau<t_1 \mid \tau \geq t_0 )
	\right)
	\\
	&\quad \quad \quad 
	+ \E\left(-\exp\left(-u(C_{t_0}) \delta t - \int_{t_1}^\tau u(C_t) \, \DT \right  ) \mid \tau\geq t_1 \right) \P( \tau \geq t_1 \mid \tau \geq t_0 )
	\Big\} \\
	&=\sup_{{C,\pi} \in {\cal A}_{w_{t_0},t_0}} \Big\{
	-  (1-s_{t_0})\exp\left(-u(C_{t_0}) \delta t \right) \\
	&\quad \quad \quad +
	s_{t_0} \exp(-u(C_{t_0}) \delta t) \E\left(-\exp\left( - \int_{t_1}^\tau u(C_t) \, \DT \right  ) \mid \tau \geq t_1 \right) \Big\} \\
	&=\sup_{{C,\pi} \in {\cal A}_{w_{t_0},t_0}} 
	\Bigg\{
	\exp\left(-u(C_{t_0}) \delta t \right) \times \\
	&\quad \quad \quad
	\left(
	-  1 +
	s_{t_0} \E\left(1-\exp\left( - \int_{t_1}^\tau u(C_t) \, \DT \right  ) \mid \tau \geq t_1 \right) \right) \Bigg\}
	\end{align*}
	The result now follows by the dynamic
	programming principle.
\end{proof}

Equation \eqref{eq:exponentialDynamicProgramming} is a one-period investment problem in a complete market.
Complete one-period markets are classified 
in \cite{armstrongClassification}. This allows us to find
a more convenient, but isomorphic, representation
of our market. 
For complete one-period markets, we may
say that two markets are isomorphic if they have the same
risk-free rate and if there is a map which acts as a 
probability space isomorphism for both the $\P$ and $\Q$
measures simultaneously.

Let $\Omega^A$ be the probability space given by $[0,1]\times[0,1]$
equipped with the Lebesgue measure. Let $q^A:[0,1]\to \R_{>0}$ be a
measurable function of integral $1$. We may define an
abstract financial market
$(\Omega^A,q^A,r)$ whose assets consist of random variables $f$
(representing the payoff of the asset)
defined on $\Omega^A$. The cost of asset $f$ is given by
\[
P^A(f):= \int_{[0,1]\times[0,1]} e^{-r \delta t} f(x,y)\,q^A(x) \, \ed x \, \ed y
\]
if this integral exists. Assets of positively infinite or undefined cost
cannot be purchased. Assets of infinitely negative cost can
be purchased at any price. The $A$ in our superscripts
stands for abstract. Notice that in this abstract market
the random variable $U$ defined by $U(x,y)=x$ is uniform
in the $\P^A$ measure and has density $q^A$ in the $\Q^A$ measure.

\begin{lemma}
	\label{lemma:bsmCanonicalForm}	
	As a one period market, the Black--Scholes--Merton market 
	from time $t_0$ to time $t_1$
	is isomorphic to the market $(\Omega^A,q^A_{\BS},r)$.
\end{lemma}
We defer the proof to appendix \ref{appendix:bsmCanonicalForm}.

Having found a simple isomorphic representative of our market, we can rewrite
the equation \eqref{eq:exponentialDynamicProgramming} in terms of the abstract market $\Omega^A$.

\begin{lemma}
	\label{lemma:reduceToCalculusVariations}
	Assume the conditions of Proposition \ref{prop:solutionExponential}.
	The value function $v_{t_0}(w_{t_0})$ can be calculated by
	solving the optimisation problem
	\begin{equation}
	\begin{aligned}
	& \underset{C \in \R, f \in L^0[0,1]}{\mathrm{maximize}}
	& & 
	\exp( -u(C)  \delta t ) \left(
	-1
	+ s_{t_0} \int_0^1 (1+v(f(s))) \, \ed s \right) \\
	& \text{subject to}
	& & C + s_{t_0} \int_0^1 e^{-r \delta t} q^{A}_{\BS}(s) f(s) \, \ed s \leq w_{t_0}.
	\end{aligned}
	\label{eq:abstractInvestmentProblem}
	\end{equation}
	taking $v=v_{t_1}$.	
\end{lemma}
\begin{proof}
	Let us write $(C_{t_0}, f)$ for a pair of a consumption
	$C_{t_0} \in \R$ and an
	investment $f \in L^0(\Omega^A)$.
	We denote by ${\cal B}_{w_{t_0}}$ the set 
	of consumptions and investments that are available with
	a budget of ${w_{t_0}}$
	\[
	{\cal B}_{w_{t_0}}=\{ (C_{t_0},f) \in \R \times L^0(\Omega^A) \mid C_{t_0} + s_{t_0} P^A_{\BS}(f) \leq w_{t_0} \}.
	\]
	If we also write
	\begin{equation*}
	\begin{split}
	U^A_{t_0}(C_{t_0},f):=
	&\exp( -u(C_{t_0})  \delta t )
	\left(-1 + s_{t_0}  \int_{[0,1]\times[0,1]} 
	\, (1+v_{t_1}( f(x,y) )) \, \ed x \, \ed y
	\right)
	\end{split}
	\end{equation*}
	to accord with equation \eqref{eq:exponentialDynamicProgramming},
	then the fact that our markets are isomorphic allows
	us to deduce that
	\begin{equation}
	v_{t_0}(w_{t_0}):=\sup_{(C,f) \in {\cal B}_{w_{t_0}}} U^A(C_{t_0},f)
	\label{eq:exponentialDynamicProgramming2}.
	\end{equation}
	
	Since $v_{t_1}$ is assumed to be concave we may average an
	investment $f(x,y)$ over the factor $y$ to obtain a new investment
	$\overline{f}$ which achieves a higher value for the gain
	function $U^A$. Thus we may restrict our attention
	to investments $f(x,y)$ which depend only upon $x$. The result follows.
\end{proof}

Note that an investment $f \in L^1$ for this abstract market
model corresponds to a derivative with payoff given by
the random variable $f(F_{\frac{\ed \Q}{\ed \P}} (\frac{\ed \Q}{\ed \P}) )$  in the original Black--Scholes--Merton market (or indeed in any isomorphic market). This derivative can then be replicated by delta hedging
in the Black--Scholes--Merton market. So the solution to the
abstract investment problem \eqref{eq:abstractInvestmentProblem}
can be straightforwardly mapped to a solution of the original problem.

\begin{lemma}
	\label{lemma:solveCalculusOfVariations}
	Assume the conditions and definitions of Proposition \ref{prop:solutionExponential}.	
	If an $\eta_{w_{t_0}}$ exists with $w^{\eta_{w_{t_0}}}=w_{t_0}$,
	then the solution of \eqref{eq:abstractInvestmentProblem} is given by $f^{\eta_{w_{t_0}}}$
	and $C^{\eta_{w_{t_0}}}$.
\end{lemma}
\begin{proof}
	We will now solve \eqref{eq:abstractInvestmentProblem} using
	the method of Lagrange multipliers. We define a vector space $V=\R \oplus L^0([0,1]) \oplus \R$
	For $\lambda \in \R$,
	we define the Lagrangian
	$L:V \to \R$ by
	\begin{equation}
	\begin{split}
	L(C,f,\lambda):=&
	\exp( -u(C)  \delta t ) \left(
	-1
	+ s_{t_0} \int_0^1 (1+v(f(s))) \, \ed s \right) \\
	& + \lambda \left( -w_{t_0} + C+  s_{t_0} \int_0^1 e^{-r \delta t} q^A_{\BS}(s) f(s) \, \ed s \right).
	\label{eq:defLagrangian}
	\end{split}
	\end{equation}
	
	Computing the directional derivatives of $L(C,f)$ we find the following
	necessary and sufficient conditions for $(C,f)$ to be a saddle point of $L(C,f,\lambda)$
	for the given $\lambda$. Firstly
	\begin{equation}
	0 \in 
	-\partial u(C) \delta t \exp( -u(C)  \delta t ) \left(
	-1
	+ s_{t_0} \int_0^1 (1+v(f(s))) \, \ed s \right) + \lambda
	\label{eq:kuhnTucker1}
	\end{equation}
	where $\partial u(C)$ is the subdifferential of $u$ at $C$.
	Secondly
	\[
	0 = \int_0^1 \left( \exp( -u(C)  \delta t ) s_{t_0} (\partial v)(f(s)) + \lambda  s_{t_0} e^{-r \delta t} q^A_{\BS}(s) \right) g(s) \, \ed s.
	\]
	The integral is well-defined since $\partial v$ will
	be single-valued almost everywhere.
	This must hold for all $g(s)$ so this is equivalent to requiring
	\begin{equation}
	(\partial v)(f(z)) =- \lambda \exp( u(C)  \delta t )  e^{-r \delta t} q^A_{\BS}(z).
	\label{eq:kuhnTucker2}
	\end{equation}
	for almost all $z\in(0,1)$.
		
	If the Kuhn--Tucker
	conditions \eqref{eq:kuhnTucker1} and \eqref{eq:kuhnTucker2}
	are satisfied, $((C,f),\lambda)$ will be a saddle point of the Lagrangian.
	The theory of Lagrange multipliers (see \cite{rockafellar} Theorem 28.3)
	now shows that if we can
	find $(C, f)$ satisfying the Kuhn--Tucker conditions \eqref{eq:kuhnTucker1}
	and \eqref{eq:kuhnTucker2} then this will yield a maximizer
	for the problem	\eqref{eq:abstractInvestmentProblem} in the case
	where the initial budget satisfies
	\begin{equation}
	w_{t_0} = C + s_{t_0} \int_0^1 e^{-r \delta t} q^A_{\BS}(s) f(s) \, \ed s.
	\label{eq:abstractProblemBudget}
	\end{equation}
	We remark that the theory of Lagrange multipliers given in \cite{rockafellar} is
	stated in terms of finite-dimensional spaces. We may, nevertheless, apply it
	by noting that if $(C, f)$ satisfies the Kuhn--Tucker conditions yet
	is not a maximizer then there must be some direction
	in which we can perturb $(C, f)$ to obtain a higher value for the gain. We may
	now apply the finite-dimensional theory to the vector space generated by this perturbation
	to obtain a contradiction.
			
	The result now follows by introducing a variable
	\[
	\eta:=-\lambda \exp( u(C) \delta t)
	\]
	to simplify the equations.
\end{proof}

This completes the proof of Proposition \ref{prop:solutionExponential}.

The outstanding difficulty is proving that an $\eta$ solving $w^\eta=w_{t_0}$ exists.
One might attempt to use general duality theory to do this. Theorem 8.3.1 of \cite{luenberger}
ensures that so long as $w_{t_0}$ is chosen to satisfy the Slater condition we can guarantee
the existence of a $\lambda$ minimizing the dual problem. However, this theorem does not
guarantee the existence of a maximizer for the primal problem. As a result, even if one knows
the value of $\lambda$ it is still unclear whether a solution to \eqref{eq:kuhnTucker1}
and \eqref{eq:kuhnTucker2} exists. When one introduces the variable $\eta$, this ensures
that $C^\eta$ and $f^\eta$ are well-defined once $\eta$ is known and so the problem
shifts to finding the correct value of $\eta$. We will resolve this issue in 
the cases of interest using a continuity argument in the next section.

\subsection{Numerical Approximation of the Multi-Period Problem}

The results of the previous section immediately suggest a numerical method for
solving our investment problems with exponential utility.

We define the minimum acceptable consumption to be
\[
C_{\min} := \inf \{ C \in \R \mid u(C)>-\infty \}.
\]
In addition to the previous assumptions that $u$ is concave and increasing, we assume
\begin{equation}
u^\dagger \text{ is continuous on } (0,\infty)
\label{eq:uDaggerCts}
\end{equation}
and
\begin{equation}
\lim_{p\to 0} u^\dagger(p)=\infty.
\label{eq:uDaggerP0}
\end{equation}
We note that our assumption that $u$ is concave and increasing also ensures that
\begin{equation}
\lim_{p\to \infty} u^\dagger(p)=C_{\min}.
\label{eq:uDaggerPInf}
\end{equation}

\begin{algoEnv}
	\label{algo:exponential}	
	Choose a grid of points
	${\cal X}=\{x_1,x_2 \ldots, x_N\}$ on which we will approximate the value function $v_t$.
	We will write $\tilde{v}_t$ for our approximate value function.
	This will be a concave increasing
	piecewise linear function equal to $-\infty$ on $(-\infty,x_1)$, linear on $[x_i,x_{i+1}]$
	and constant on $[x_N,\infty)$. We will simply need
	to store the values ${\tilde{v}}_t(x_i)$ at the grid points.	
	
	To avoid numerical overflow issues we define a function
	$\ell(x):=-\log(-x)$ and store the values $\ell(\tilde{v}_t(x_i))$
	at each grid point rather than storing $\tilde{v}_t(x_i)$ itself.
	\begin{enumerate}[label=(\roman*)]
		\item Choose the values at the final time point $T-\delta t$ by
		\[
		\tilde{v}_{T-\delta t}( x_i ) := v_{T-\delta t}(x_i) = -\exp(-u(x_i) \delta t).
		\]
		Or equivalently
		\[
		\ell(\tilde{v}_{T-\delta t}( x_i )) = \ell(v_{T-\delta t}(x_i)) = u(x_i) \delta t.
		\]
		\item Suppose that $\tilde{v}_t$ is known. Set
		$\tilde{v}_{t-\delta t}(x_i)$ to be the solution of \eqref{eq:abstractInvestmentProblem}
		with $v_{t_1}=\tilde{v}_t$ and initial budget $x_i$. We describe in detail how
		to solve this problem in Proposition \ref{prop:logFormaule} below.
	\end{enumerate}
\end{algoEnv}

Since $v_{T-\delta t}$ is concave and increasing and
$\tilde{v}_{T-\delta t}$ is piecewise linear $\tilde{v}_{T-\delta t}(w)\leq v_{T-\delta t}(w)$. Let $\hat{v}_{t}(w)$
be defined to be the solution of \eqref{eq:abstractInvestmentProblem}
with $v_{t_1}=\tilde{v}_t$ and initial budget $w$. We see
that $\tilde{v}_{T-\delta t}(w)\leq \hat{v}_{T-\delta t}(w)
\leq v_{T-\delta t}(w)$.

Let
${\cal X}_1 \subseteq {\cal X}_2 \subseteq {\cal X}_3 \ldots $ 
be an increasing sequence of grids
with ${\cal X}_\infty:=\cup_{j=1}^\infty {\cal X}_j$ being dense
in $(0,\infty)$. Write $\tilde{v}^j_t$
for the approximations with respect to ${\cal X}_j$.
We see by repeating the argument above that $\tilde{v}^j_t(w)\leq v_t(w)$ at all points $w \in (0,\infty)$.
Hence we may define
\[
\tilde{v}_t(w)=\lim_{j\to \infty} \tilde{v}^j_t(w).
\]

\begin{theorem}[Convergence of Algorithm \ref{algo:exponential}]
	Define
	\[
	W_{\min,t} = \sup\{ w \mid v_t(w) = -\infty \}.
	\]
	For $w > W_{\min,t}$ we have
	\[
	\tilde{v}_t(w) = v_t(w).
	\]
\end{theorem}
\begin{proof}
	Let ${\cal V}$ denote the space of concave, increasing functions $v(w)$ which satisfy $v(w)=-\infty$ for $w<0$ and where $v(w)$ is bounded above by $0$.
	For two adjacent times $t_0, t_1=t_0+\delta t$ in our grid 
	we define a solution function $\phi_{t_0,t_1,w}:{\cal V} \to \R$ by
	setting
	\[
	\phi_{t_0,t_1,w}(v_{t_1})
	\]
	to equal the supremum in \eqref{eq:exponentialDynamicProgramming}. By composing these
	solution functions in the obvious way, we obtain a solution function $\phi_{t_0,t_1,{w}}$
	for any times in the grid with $t_0\leq t_1$.
	
	We define a corresponding minimum budget as follows:
	\[
	W_{\min,t_0,t_1}(v) = \sup\{ w \mid \phi_{t_0,t_1,w}(v) = -\infty \}.
	\]
	
	Let $t_0, t_1$ be adjacent times in the grid.
	Given $v \in {\cal V}$ with $\phi_{t_0,t_1,w}(v)$ finite,
	let $(C_{t_0}, \pi) \in {\cal A}_{w,t_0,t_1}$
	be a maximizing strategy for the problem \eqref{eq:exponentialDynamicProgramming} with $v_{t_1}=v$.
	Suppose $v^\prime \in {\cal V}$. We have
	\begin{multline*}
	\Big|
	\exp\left(-u(C_{t_0}) \delta t \right)
	\left( -1 + s_{t_0} 
	\E\left( 1+v(w^{(C_{t_0},\pi)}_1) \right)
	\right)
	- \\
	\exp\left(-u(C_{t_0}) \delta t \right)
	\left( -1 + s_{t_0} 
	\E\left( 1+v^\prime(w^{(C_{t_0},\pi)}_1) \right)
	\right)
	\Big| \\
	\leq A \exp(-u(C_{t_0}) \delta t) \|v-v^\prime\|_\infty
	\end{multline*}
    for some constant $A$.
	Hence for any $\epsilon>0$ we can find $\delta_1>0$ such that $\|v-v^\prime\|_\infty<\delta_1$ implies
	\[
	\phi_{t_0,t_1,w}(v^\prime) \geq \phi_{t_0,t_1,w}(v)-\epsilon.
	\]
	We have shown $\phi_{t_0,t_1,w}$ is lower semi-continuous in the $\sup$ norm for adjacent
	times $t_0$ and $t_1$. It follows that $\phi_{t_0,t_1,w}$ is lower semi-continuous for all $t_0<t_1$.
	
	Given $v \in {\cal V}$ and $h \in \R$, define the translation
	\[
	v_h(x)=\begin{cases}
	v(x-h) & x- h \geq 0 \\
	-\infty & x - h < 0.
	\end{cases} = \min \{ v(x-h), (\sup v) \id_{x-h<0} \}
	\]
	Define $f_{t_0,t_1,w,v}(h)=\phi_{t_0,t_1,w}(v_h)$. The function $v(x,h)=v_h(x)$
	is concave.
	Hence $f_{t_0,t_1,w,v}$
	is concave as a function of $h$. If $w>W_{\min,t_0,t_1}(v)$ then 
	$0 \in \ri(\dom f_{t_0,t_1,w,v})$, where $\ri f$ denotes the relative interior of $f$. Hence
	$f_{t_0,t_1,w,v}$ is continuous in $h$ at $0$. 
	
	Combining this with the lower semi-continuity result, we see that if $w>W_{\min,t_0,t_1}(v)$
	then given $\epsilon>0$, we can find $\delta_1>0$ and $\delta_2>0$ such that
	\[
	\phi_{t_0,t_1,w}(v_{\delta_1}(x) - \delta_2) \geq 
	\phi_{t_0,t_1,w}(v) - \epsilon.
	\]
	Let us write $v_\epsilon(x)$ for the function $v_{\delta_1}(x)-\delta_2$. Given a function $f$
	let us write $\Gamma_f$ for the {\em hypograph} of $f$, that is to say the set of points on or below the graph. We have $\Gamma_v \supseteq \Gamma_{v_\epsilon}$. For any function $v^\prime \in {\cal V}$ satisfying
	$\Gamma_v \supseteq \Gamma_{v^\prime} \supseteq \Gamma_{v_\epsilon}$ we will have
	\[
	\phi_{t_0,t_1,w}(v) \geq \phi_{t_0,t_1,w}(v^\prime) \geq \phi_{t_0,t_1,w}(v_\epsilon)
	\geq 
	\phi_{t_0,t_1,w}(v) - \epsilon.
	\]
	since it is clear that $\Gamma_{v} \supseteq \Gamma_{v^\prime}$ implies $\phi_{t_0,t_1,w}(v) \geq \phi_{t_0,t_1,w}(v^\prime)$. Note that we can always find a piecewise linear approximation
	between $\Gamma_v$ and $\Gamma_{v_\epsilon}$.
	
	Given a value for $\epsilon_0$, we may inductively extend this to a sequence of
	positive ${\epsilon_t}$ for $t \in {\cal T}$
	such that if our approximation $\tilde{v}_t$ satisfies
	$\Gamma_{v_t} \supseteq \Gamma_{\tilde{v}_t} \supseteq \Gamma_{(v_t)_{\epsilon_t}}$
	then it will automatically satisfy 
	$\Gamma_{v_{t-\delta t}} \supseteq \Gamma_{\tilde{v}_{t-\delta t}} \supseteq \Gamma_{(v_{t - \delta t})_{\epsilon_{t-\delta t}}}$.
	By choosing a sufficiently fine grid we can ensure this condition is satisfied at time $T-\delta t$.
	By further refinements we may ensure that it is satisfied at all times.
\end{proof}

Let us now describe in full detail how to
solve \eqref{eq:abstractInvestmentProblem}
given that $v_{t_1}$ is of the form used in our algorithm.
In Proposition \ref{prop:logFormaule}, we will give the formulae
necessary to solve the problem on a computer in a format that addresses
numerical overflow issues.  Terms on the left hand side of the equations in the Proposition
should be stored in computer memory and can be computed without overflow issues from the terms on the right. We use infinite values for some terms as a convenient shorthand, terms such as an exponential of $-\infty$ should be interpreted in the obvious way.

To store probability values we define a bijection $L:[0,1]\to \R\cup \{\pm \infty \}$
by
\[
L(u) = \begin{cases}
\log(2u ) & u\leq0.5 \\
-\log(2-2u) & u> 0.5.
\end{cases}
\]
We note that the GNU scientific library contains a function {\tt gsl\_sf\_log\_erfc}
which computes the logarithm of the complementary error function which we can then use to compute $L(\Phi)$.

We define a function
\[
\tilde{u}(y) = \log( u^\dagger (e^y) ).
\]
For the specific functional form
\[
u(x)=\begin{cases}
a (x-w_{t_0})^n + b & x \geq 0 \\
-\infty & \text{otherwise}.
\end{cases}
\]
We may compute
$\tilde{u}$ without experiencing
overflow errors using the formulae
\begin{align}
\tilde{u}_0(p)&:= \frac{1}{n - 1}(p - \log(a \, n)), \\
\tilde{u}(y)&= \begin{cases}
\log( e^{\tilde{u}_0(p)} ) & w_{t_0} = 0 \\
\log( e^{\tilde{u}_0(p)} + e^{\log(w_{t_0})}) & w_{t_0} > 0 \\
\log( e^{\tilde{u}_0(p)} - e^{\log(-w_{t_0})} ) & \tilde{u}_0(p) > \log(-w_{t_0}) \text{ and } w_{t_0} < 0 \\
-\infty & \tilde{u}_0(p) \leq \log(-w_{t_0}) \text{ and } w_{t_0} < 0.
\end{cases}
\end{align}
We note the standard approach to computing the log of sums and differences of exponentials
without overflow issues should be used when evaluating
expressions such as this.

\begin{proposition}
	\label{prop:logFormaule}
	Let $v$ be a concave, non-positive, increasing function
	which is linear between grid points	in ${\cal X}=\{x_1,x_2, \ldots x_N\}$
	with $x_i$ strictly increasing. Suppose also that $v$ is equal
	to $-\infty$ on $(-\infty,x_1)$ and constant on $(x_N,\infty)$.
    Suppose that $u^\dagger$ is continuous and 
	satisfies equations \eqref{eq:uDaggerCts} and \eqref{eq:uDaggerP0}.
	
	Define a decreasing sequence of points $\log( p_i )$ by
	\begin{equation}
	\log( p_i ) = \begin{cases}
	\infty & i=0 \\
	\log (  e^{ - \ell( v(x_{i})) } -e^{-\ell(v(x_{i+1}))} ) - \log(x_{i+1}-x_{i}) & 0<i<N \\
	-\infty & i=N.
	\end{cases}
	\label{eq:defLogDualPoints}
	\end{equation}
	For a given value of $\log \eta$, define $L(U^\eta_i)$ and $L(Q^\eta_i)$ for $0<i<N$ by
	\begin{align}
	L(U^\eta_i) &= L \left( \Phi\left( -\frac{1}{2}M + \frac{1}{M} \left( \log (\eta ) - r \delta t  - \log( p_i ) \right) \right) \right),
	\label{eq:logUIExplicit} \\
	L(Q^\eta_i) &= L \left( \Phi\left( \frac{1}{2}M + \frac{1}{M} \left( \log (\eta ) - r \delta t - \log( p_i ) \right) \right) \right).
	\label{eq:logQIExplicit}
	\end{align}
	Define $L(U^\eta_0)=L(Q^\eta_0)=-\infty$ and $L(U^\eta_N)=L(Q^\eta_N)=\infty$.
	We may then define the quantity $A^\eta$ by
	\[
	A^\eta
	=\log\left(
	e^{\log(1-s_{t_0})}+ \sum_{i=1}^N e^{ \log(s_{t_0}) + \log(-v(x_i)) + \log\left(e^{\log U^\eta_{i}}-e^{\log U^\eta_{i-1}} \right) }\right).
	\]
	We then have that
	\begin{equation}
	\log (C^\eta) = \tilde{u}( \log(\eta) - \log(\delta t) - A^\eta )
	\label{eq:logGammaEtaExplicit}
	\end{equation}
	where $C^\eta$ is as defined in \eqref{eq:defCEta}.
	We have
	\begin{equation}
	\log(w^\eta)= \log\left(
	e^{\log( C^\eta)} +  \sum_{i=1}^N e^{ \log(s_{t_0}) -r \delta t + \log(x_i)+\log\left(e^{\log Q^\eta_i}-e^{\log Q^\eta_{i-1}} \right)}
	\right)
	\label{eq:logxEtaExplicit}
	\end{equation}
	and $w^\eta$ depends continuously upon $\eta$.
	If $w_{t_0} > s_{t_0} e^{-r \delta t} x_1 + C_{\min}$,
	we may find the value of $\eta_{w_{t_0}}$ by finding
	$\log(\eta)$ such that $\log(w^\eta)=\log(w_{t_0})$.
	We then have
	\begin{align}
	\ell( v_{t_0}(w_{t_0}))=
	u(C^\eta)  \delta t - A^\eta.
	\label{eq:logV0Explicit}
	\end{align}
	If $w_{t_0} < s_{t_0} e^{-r \delta t} x_1$, the maximum in \eqref{eq:abstractInvestmentProblem} is $-\infty$ which is achieved by the negative consumption $C = w_{t_0} - s_{t_0} e^{-r \delta t} x_1$.
\end{proposition}
\begin{proof}
	Corresponding to \eqref{eq:defLogDualPoints} we have a decreasing sequence of points $p_i$ given
	by
	\begin{equation}
	p_i = \begin{cases}
	\infty & i=0 \\
	\frac{v(x_{i+1})-v(x_{i})}{x_{i+1}-x_{i}} & 0<i<N \\
	0 & i=N.
	\end{cases}
	\label{eq:defDualPoints}
	\end{equation}
	We will then have
	\[
	v^\dagger(p) =
	\sum_{i=1}^N x_i \id_{[p_{i},p_{i-1})}(p).
	\]
	From \eqref{eq:defFEta}
	\[
	f^\eta(u) = \sum_{i=1}^N x_i \id_{[p_{i},p_{i-1})}\left( 
	\eta  e^{-r \delta t} q^A_{\BS}(u) \right).
	\]
	Hence we will be able to deduce that
	\begin{equation}
	f^\eta(U) = \sum_{i=1}^N x_i \id_{(U^\eta_{i-1},U^\eta_i]}\left( 
	U \right)
	\label{eq:fEtaAsSum}
	\end{equation}
	if we can show \eqref{eq:logUIExplicit} ensures that
	\begin{equation}
	\eta e^{-r \delta t} q^A_{\BS}(U^\eta_i) = p_i.
	\label{eq:uIImplicit}
	\end{equation}
	Writing $\phi$ for the pdf of the standard normal we compute
	\begin{align*}
	q^A_{\BS}(u) &=
	\frac{\phi( M + \Phi^{-1}(u))}{\phi( \Phi^{-1}(u))} \\
	&= \exp\left( \frac{1}{2}( \Phi^{-1}(u)^2 - (M + \Phi^{-1}(u))^2 )\right) \\
	&= \exp\left( -\frac{1}{2}M^2 - M \Phi^{-1}(u) \right).
	\end{align*}
	Hence equation \eqref{eq:uIImplicit} is equivalent to
	\begin{equation}
	U^\eta_i = \Phi\left( -\frac{1}{2}M - \frac{1}{M} \log \left( \frac{1}{\eta} e^{r \delta t} p_i \right) \right).
	\label{eq:uIExplicit}
	\end{equation}
	for $0<i<N$, which will hold due to our definition \eqref{eq:logUIExplicit}.
	From \eqref{eq:defCEta} and \eqref{eq:fEtaAsSum} we have
	\begin{align}
	C^\eta  &= u^\dagger \left( -\frac{\eta}{\delta t}
	\left(
	-1
	+ s_{t_0} \int_0^1 \left(1+v\left( \sum_{i=1}^N x_i \id_{(U^\eta_{i-1},U^\eta_i]}(s) \right) \right) \, \ed s \right)^{-1}
	\right) \nonumber \\
	&= u^\dagger \left( -\frac{\eta}{\delta t}
	\left(
	-1
	+ s_{t_0} \int_0^1 \left(1+ \sum_{i=1}^N v(x_i) \id_{(U^\eta_{i-1},U^\eta_i]}(s) \right) \, \ed s \right)^{-1}
	\right) \nonumber \\
	&= u^\dagger \left( -\frac{\eta}{\delta t}
	\left(
	-1
	+ s_{t_0} \left(1+ \sum_{i=1}^N v(x_i) (U^\eta_i-U^\eta_{i-1})\right) \right)^{-1}
	\right).
	\end{align}
	Equation \eqref{eq:logGammaEtaExplicit} follows immediately.
	
	Use \eqref{eq:defWeta} and \eqref{eq:defQ} to see that
	\begin{align}
	w^\eta &= C^\eta + s_{t_0} \sum_{i=1}^N \int_{U^\eta_{i-1}}^{U^\eta_i} e^{-r \delta t} q^A_{\BS}(s) x_i \, \ed s \nonumber \\
	&= C^\eta + s_{t_0} \sum_{i=1}^N e^{-r \delta t} x_i (Q(U^\eta_i)-Q(U^\eta_{i-1})) \nonumber \\
	&= C^\eta + s_{t_0} \sum_{i=1}^N e^{-r \delta t} x_i (Q^\eta_i-Q^\eta_{i-1})
	\label{eq:xEtaExplicit}
	\end{align}
	The last line follows directly from our definitions of $Q$, $U^\eta_i$ and $Q^\eta_i$.
	We now see that equation \eqref{eq:xEtaExplicit} is equivalent to
	\eqref{eq:logxEtaExplicit}.
	
	Our explicit formula, \eqref{eq:uIExplicit}, for $U^\eta_i$ shows that it depends continuously on $\eta$	given the assumption \eqref{eq:uDaggerCts}.
	It then follows from equation \eqref{eq:xEtaExplicit} that $w^\eta$ depends continuously on $\eta$.
	Lemmas \ref{lemma:xForSmallEta} and \ref{lemma:xForLargeEta} below then establish
	that we can solve for $\eta$ in $w^\eta = w_{t_0}$ under the conditions of the proposition.
	
	The value function
	is then given by
	\begin{align*}
	v(t_0,w^\eta)&=
	\exp( -u(C^\eta)  \delta t ) \left(
	-1
	+ s_{t_0} \int_0^1 (1+v( 
	\sum_{i=1}^N x_i \id_{(U^\eta_{i-1},U^\eta_i]}\left( 
	s \right)
	)) \, \ed s \right) \\
	&=
	\exp( -u(C^\eta)  \delta t ) \left(
	-1
	+ s_{t_0}(1+ 
	\sum_{i=1}^N v(x_i) (U^\eta_{i}-U^\eta_{i-1}) ) \right)
	\end{align*}
	and so \eqref{eq:logV0Explicit} also follows.
\end{proof}

\begin{lemma}
\label{lemma:xForSmallEta}	
Under the assumptions of Proposition \ref{prop:logFormaule},
\[
\lim_{\eta \to 0} w^\eta = \infty.
\]	
\end{lemma}
\begin{proof}
Our assumptions on $v$ ensure that
\[
-1 +s_{t_0} \int_0^1 (1 + v(f^\eta(s))) \, \ed s \leq -1+s_{t_0} <0.
\]	
Hence
\[
\frac{1}{-1+s_{t_0}} < \left( -1 +s_{t_0} \int_0^1 (1 + v(f^\eta(s))) \right)^{-1} < 0.
\]
It now follows from our equation \eqref{eq:uDaggerP0} coupled with equation \eqref{eq:defCEta}
that
\[
\lim_{\eta \to 0} C^\eta = \infty.
\]
The result now follows from \eqref{eq:defWeta}.
\end{proof}
\begin{lemma}
Under the assumptions of Proposition \ref{prop:logFormaule}, 
\[
\lim_{\eta \to \infty} C^\eta = C_{\min}.
\]	
\label{lemma:limGammaEta}
\end{lemma}	
\begin{proof}
Our assumptions on $v$ ensure that
\[
\left(
-1
+ s_{t_0} \int_0^1 (1+v(f^\eta(s))) \, \ed s \right)^{-1}
\]
is bounded. Hence using the expression \eqref{eq:defCEta}
combined with assumption \eqref{eq:uDaggerPInf} we find $C^\eta \to C_\text{min}$
as $\eta \to \infty$.
\end{proof}
\begin{lemma}
Under the assumptions of Proposition \ref{prop:logFormaule}, 
\[
\lim_{\eta \to \infty} w^\eta = C_{\min} + s_{t_0} e^{-r \delta t } x_1 .
\]		
\label{lemma:xForLargeEta}	
\end{lemma}
\begin{proof}
Define 
\[
p^* = \inf \partial v(x_1).
\]	
For $\eta >0$, define
\begin{equation}
s^*_\eta = (q^A_{\BS})^{-1}(p^* \eta^{-1} e^{r\delta t}),
\label{def:sStarEta}
\end{equation}
which ensures that
\begin{equation}
s \geq s_\eta^* \iff \eta e^{-r\delta t} q^A_{\BS}(s) < p^{\star}.
\label{eq:sStarCondition}
\end{equation}
We compute
\begin{align}
\int_0^1 q^A_{\BS}(s) f^\eta(s) \ed s 
&= \int_0^1 q^A_{\BS}(s) v^\dagger( \eta  e^{-r\delta t} q^A_{\BS}(s) ) \, \ed s
\nonumber \\ 
&= 
\int_0^{s^*_\eta} q^A_{\BS}(s) v^\dagger( \eta  e^{-r\delta t} q^A_{\BS}(s) ) \, \ed s
\nonumber \\
&\qquad + \frac{1}{\eta e^{-r\delta t}} \int_{s^*_\eta}^1 \eta  e^{-r\delta t} \, q^A_{\BS}(s) v^\dagger( \eta  e^{-r\delta t} q^A_{\BS}(s) ) \, \ed s
\nonumber \\
&\leq 
\int_0^{s^*_\eta} q^A_{\BS}(s) x_1 \, \ed s
\nonumber \\
&\qquad + \frac{1}{\eta  e^{-r\delta t}} \int_{s^*_\eta}^1 \eta e^{-r\delta t} \, q^A_{\BS}(s) v^\dagger( \eta e^{-r\delta t} q^A_{\BS}(s) ) \, \ed s 
\label{eq:xEtaInequality1}
\end{align}

We note that $p \in \partial v( v^\dagger(p))$.
By the definition of the subdifferential at $v^\dagger(p)$
\[
v(x) \leq v(v^\dagger(p)) + p( x - v^\dagger(p)).
\]
Rearranging yields
\[
p v^\dagger(p) \leq p x + v(v^\dagger(p)) - v(x).
\]
Using the fact $v$ is increasing and substituting $x_1$ for $x$ we find that for all $p$
\[
p v^\dagger(p) \leq p x_1 + v(x_N) - v(x_1).
\]
Using this inequality in \eqref{eq:xEtaInequality1}  we find
\begin{align}
\int_0^1 q^A_{\BS}(s) f^\eta(s) \ed s 
&\leq 
\int_0^{s^*_\eta} q^A_{\BS}(s) x_1 \, \ed s 
\nonumber \\
&\qquad + 
\frac{1}{\eta  e^{-r\delta t}}
\int_{s^*_\eta}^1 ( \eta  e^{-r\delta t} \, q^A_{\BS}(s) x_1 + v(x_N) - v(x_1)) \, \ed s
\nonumber \\
&\leq 
\int_0^{s^*_\eta} q^A_{\BS}(s) x_1 \, \ed s
\nonumber \\
&\qquad + 
\frac{1}{\eta e^{-r\delta t}}
\int_{s^*_\eta}^1 ( p^* x_1 + v(x_N) - v(x_1)) \, \ed s.
\label{eq:xEtaInequality2}
\end{align}
by \eqref{eq:sStarCondition}. From \eqref{def:sStarEta}
\[
\lim_{\eta \to \infty} s^*_\eta = 1.
\]
We may therefore take the limit of the inequality \eqref{eq:xEtaInequality2}
to find
\[
\liminf_{\eta \to \infty}
\int_0^1 q^A_{\BS}(s) f^\eta(s) \ed s  \leq x_1.
\]
Using this, Lemma \ref{lemma:limGammaEta} and the definition of $w^\eta$ in
equation \eqref{eq:defWeta} we find
\[
\liminf_{\eta \to \infty} w^\eta \leq C_{\min} + s_{t_0} e^{-r \delta t} x_1.
\]
From \eqref{eq:defCEta} and \eqref{eq:defWeta} one sees that, on the other hand, 
for all $\eta>0$ we have
\[
w^\eta \geq C_{\min} + s_{t_0} e^{-r \delta t} x_1.
\]
The result follows.
\end{proof}

\begin{remark}
	We note that if we follow the optimal investment
	strategy at time $t$, then the optimal investment strategy will result
	in a wealth at time $t+\delta t$ which takes values in the grid $\{x_1,\ldots,x_n\}$.
	We may then approximate the value function on the space-time grid
	$\{x_1,\ldots x_n \} \times \{0,\delta t, 2 \delta t, \ldots, T\}$.
	One can then obtain a simulation of the optimal strategy by first simulating 
	the stock price on the time grid and then computing the corresponding dynamics
	of $x_t$ in the grid $\{x_1,\ldots x_n \}$ using this approximation to the value
	function. Since the wealth process never leaves a fixed space-time grid, we can use
	the same approximation of the value function for all the scenarios.
\end{remark}

\begin{remark}
	When implementing this algorithm we notice that many values
	of $U_i^\eta$ will be extremely close to either 0 or 1, and so including
	these terms will have a negligible effect on the values of the sums
	in the equations \eqref{eq:logGammaEtaExplicit}, \eqref{eq:logxEtaExplicit}.
	Financially this is equivalent to ignoring extreme events of
	very low probability where the
	$\P$ and $\Q$ disagree by a large amount. Since our payoff functions $f$
	take values in ${\cal X}$, and so are bounded and positive, ignoring
	these extreme events will have no material impact upon either the price or the expected utility.
	The value we chose in our numerical calculations was $\epsilon=10^{-10} \max{|v(x_i)|}^{-1}$.
	
	This can be used to speed up the algorithm. When
	calculating $w^\eta$, choose some small $\epsilon$ and define
	\begin{align*}	
	i_{\min}&:=\max \{1\} \cup \{ i \mid U_i<\epsilon\} \\
	i_{\max}&:=\min  \{N\} \cup \{ i \mid U_i>1-\epsilon\}.
	\end{align*}
	To compute these values and the values of $U_i$,
	first use the method of bisection to find some
	$i^*$ where $\epsilon<U_{i^*}<1-\epsilon$. Then compute the values of $U_i$
	from $i^*$ down to $i_{\min}$, stopping when $U_i<\epsilon$. Similarly compute the values of $U_i$ from $i^*$ up to $i_{\max}$, stopping when $U_i>1-\epsilon$.
	No other values of $U_i$ outside the range $i_{\min-1}\leq i \leq i_{\max}$ are then needed in the computation of $w^\eta$.
	
	When computing the values of the sums in \eqref{eq:logGammaEtaExplicit}, \eqref{eq:logxEtaExplicit} use indices running from $i_{\min}$ to $i_{\max}$ rather than from $1$ to $n$. 
\end{remark}

\section{Proof of Lemma \ref{lemma:bsmCanonicalForm}}
\label{appendix:bsmCanonicalForm}

\begin{proof}
	
	If $\mu=r$, then the result is trivial. We will consider
	the case $\mu>r$, the case $\mu<r$ is similar.
	
	The classification of complete markets already shows that
	the Black--Scholes--Merton market over the time period $[t_0,t_1]$
	is isomorphic to a
	market of this form for an appropriate choice of $q^A$
	which we will call $q^A_{\BS}$. Let
	$\frac{\ed \Q}{\ed \P}$ denote the Radon--Nikodym derivative
	of the measures $\Q$ and $\P$ in the Black--Scholes--Merton market.
	Let $F_\frac{\ed \Q}{\ed \P}$ denote the $\P$-measure distribution
	function of the Radon--Nikodym derivative.
	The classification theorem moreover gives us
	an isomorphism
	for both the $\P$ and $\Q$ measures which maps the uniformly distributed random variable $U^\prime:=F_\frac{\ed \Q}{\ed \P}(\frac{\ed Q}{\ed P})$
	to $U$. In particular this tells us that 
	\begin{equation}
	\int_{0}^{w} q^A_{\BS}(s) \ed s = \P_{\Q^A}(U \leq w)
	= \P_{\Q}(F_\frac{\ed \Q}{\ed \P}(U^\prime \leq w) )
	\label{eq:defQInProof}
	\end{equation}
	Differentiating this, we may obtain an expression for $q^A_{\BS}$.		
	
	The $\P$ measure distribution function of the log stock price, $z_{t_1}=\log(S_{t_1})$ given the log stock price $z_{t_0}$
	in the Black--Scholes--Merton model is
	\[
	p(z) = \frac{1}{\sqrt{2 \pi \sigma^2 \delta t }}
	\exp\left( -\frac{(z-(z_{t_0}+(\mu-\frac{1}{2}\sigma^2)\delta t))^2}{2 \sigma^2 \delta t}
	\right).
	\]
	Similarly the
	$\Q$ measure distribution function of $z_{t_1}$ is
	\[
	q(z) = \frac{1}{\sqrt{2 \pi \sigma^2 \delta t }}
	\exp\left( -\frac{(z-(z_{t_0}+(r-\frac{1}{2}\sigma^2)\delta t))^2}{2 \sigma^2 \delta t}
	\right).
	\]
	The standard computation of the $\Q$ measure using Girsanov's theorem
	shows that
	\begin{align*}
	\frac{\ed \Q}{\ed \P}(z)&=\frac{q(z)}{p(z)}.
	\end{align*}
	Hence
	\begin{align*}
	\frac{\ed \Q}{\ed \P}(z) &= \exp\left( -\frac{(z-(z_{t_0}+(r-\frac{1}{2}\sigma^2)\delta t))^2
		- (z-(z_{t_0}+(\mu-\frac{1}{2}\sigma^2)\delta t))^2}{2 \sigma^2 \delta t}
	\right).
	\end{align*}
	Note that the term inside the $\exp$ is linear in $z$, so $\frac{\ed \Q}{\ed \P}$
	is decreasing. Hence
	$U^\prime(z)$ is decreasing, and we recall that $U^\prime$ is uniformly distributed. Hence, $U^\prime(z)=1-F_z(z)$ where $F_z$ is the $\P$-measure
	distribution function of $z_t$. But conditioned on $z_{t_0}$, $z_{t_1}$ is normally distributed with mean $\mu-\tfrac{1}{2}\sigma^2$ and
	standard deviation $\sigma \sqrt{\delta t}$. Hence
	\[
	z_{t_1} = z_{t_0} + (\mu-\tfrac{1}{2}\sigma^2) \delta t + 
	\sigma \sqrt{\delta t} \, \Phi^{-1}(U^\prime)
	\]
	where $\Phi$ is the cumulative distribution function of the standard normal
	distribution.
	
	We now compute
	\begin{align*}
	\P_\Q( U^\prime \leq w )
	&=\P_\Q( z_{t_1} \leq
	z_{t_0}+(\mu-\tfrac{1}{2}\sigma^2)\delta t + 
	\sigma \sqrt{\delta t} \Phi^{-1}( w ) ) \\
	&=\P_\Q( z_{t_1} \leq
	z_{t_0}+(r-\tfrac{1}{2}\sigma^2)\delta t
	+ (\mu-r) \delta t
	+
	\sigma \sqrt{\delta t} \Phi^{-1}( w ) ).
	\end{align*}
	Since $z_{t_1}$ is normally distributed in the $\Q$ measure with 
	mean $r-\tfrac{1}{2}\sigma^2$ and standard deviation $\sigma \sqrt{\delta t}$
	we find
	\[
	\P_\Q( U^\prime \leq w ) = \Phi\left( \left| \frac{(\mu-r)\sqrt{\delta t}}{\sigma}
	\right|  + \Phi^{-1}(w) \right).
	\]
	Combining this with \eqref{eq:defQInProof}, we get the result.
\end{proof}